\documentclass[aps,pra,twocolumn,showpacs,amsmath,amstex,amssymb,citeautoscript]{revtex4-2}
\usepackage{physics}
\usepackage{siunitx}
\usepackage[colorlinks,linkcolor=magenta,citecolor=magenta]{hyperref}
\usepackage{bm}
\usepackage{graphicx}

\usepackage{tikz}

\newcommand*\Circled[1]{\tikz[baseline=(char.base)]{
            \node[shape=circle,draw,inner sep=1pt] (char) {#1};}}

\usepackage{xcolor}
\usepackage{lipsum}

\begin{document}

\title{Complex-valued 3D atomic spectroscopy with Gaussian-assisted inline holography}

\author{Xing Huang$^1$}
\email{xinghuang18@fudan.edu.cn}
\author{Yuzhuo Wang$^{1,2}$}
\author{Jian Zhao$^1$}
\author{Saijun Wu$^1$}
\email{saijunwu@fudan.edu.cn}

\address{$^1$Department of Physics, State Key Laboratory of Surface Physics and Key Laboratory of Micro and Nano Photonic Structures (Ministry of Education), Fudan University, Shanghai 200433, China.\\ $^2$State Key Laboratory of Quantum Optics and Quantum Optics Devices, Institute of Laser Spectroscopy,
Shanxi University, Taiyuan 030006, China.
}




\begin{abstract}
When a laser-cooled atomic sample is optically excited, the envelope of coherent forward scattering can often be decomposed into a few complex Gaussian profiles. The convenience of Gaussian propagation helps addressing key challenges in digital holography. In this work, we develop a Gaussian-decomposition-assisted approach to inline holography, for single-shot, simultaneous measurements of absorption and phase-shift profiles of small atomic samples sparsely distributed in 3D. 
The samples’ axial positions are resolved with micrometer resolution, and their spectroscopy are extracted from complex-valued images recorded at various probe frequencies. The phase-angle readout is not only robust against transition saturation, but also insensitive to atom-number and optical-pumping-induced interaction-strength fluctuations. Benefiting from such features, we achieve hundred-kHz-level single-shot resolution to the transition frequency of a $^{87}$Rb D2 line, with merely hundreds of atoms. We further demonstrate single-shot 3D field sensing by measuring local light shifts to the atomic array with micrometer spatial resolution.

\end{abstract}

\maketitle

\section{Introduction}\label{sec:intro}

Atoms are ideal quantum sensors. By measuring the optical response of free atoms, the centers and linewidths of atomic levels can be accurately inferred for unveiling unknown interactions and to calibrate external potentials~\cite{Sansonetti2011,Lu2013,Beyer2017, Fuchs2018}. The optical transition properties are affected by the center-of-mass motion. The laser-cooling techniques, which are able to effectively freeze the atomic motion, were expected to greatly enhance the precision of atomic spectroscopy~\cite{MetcalfBook}. Nowadays, however, while laser cooling and trapping are often essential in spectroscopic measurements on narrow transitions~\cite{Wynands2005, Ludlow2015, Safronova2018, Rengelink2018, Asenbaum2020}, precision spectroscopy on strong transitions tends to rely more on traditional sources such as saturated vapors~\cite{Starkind2023} and atomic beams~\cite{Zheng2017}. Compared to the thermal sources, the fluxes from cold-atom sources are moderate and easily fluctuate.
Furthermore, during a prolonged probe period, the  light-atom interaction strength, characterized by the atomic polarizability $\alpha$, can be modified by optical pumping even at extremely low light levels~\cite{Brown2013a}. Efficient suppression of the systematic errors associated with such fluctuations are important for atomic precision spectroscopy to fully benefit from the long-time Doppler-free interaction offered by laser cooling, and to enable novel applications of ultra-cold atoms such as for wideband quantum sensing of 3D potentials
~\cite{Peyrot2019a, stamperkurn,Hilton2020}.  

The impacts of atom number and interaction-strength 
fluctuations can be substantially suppressed in atomic spectroscopy, if one is able to extend the spectroscopic data from real to complex numbers. In particular, for a dilute sample subjected to a near-resonant probe~\cite{Hung2011}, both the optical depth ${\rm OD}$ and phase shift $\phi$ are proportional to the atom number $N$ and the modulus of the atomic polarizability $|\alpha|$. But the phase angle $\beta={\rm arg}(\phi+i{\rm OD}/2)$ is largely decided by the atomic phase angle, ${\rm arg}(\alpha)$, highly insensitive to the $N|\alpha|$ fluctuations~\cite{Wang2022b,Zhao2022b}. 
Furthermore, as being discussed in our recent work, the $\beta$-value is robust against power broadening effects~\cite{Zhao2022b} to support efficient readouts. 
Unfortunately, it is not possible to simultaneously measure ${\rm OD}$ and $\phi$ distributions without precisely knowing the complex $E_r'$ and $E_r$, the transmitted probe wavefronts in presence and absence of the atomic sample respectively. In standard transmission imaging techniques where the light intensities are recorded only~\cite{Hung2011,Higbie2005}, the optical phase information are completely lost. While single-mode coherent spectroscopy~\cite{Aljunid2009,Kohnen2011, Pototschnig2011, Fischer2017} is able to retrieve the $E_r$-mode-averaged ${\rm OD}$ and $\phi$, the single-mode readouts do not support spatial resolution. Furthermore, since details of atomic distribution are disregarded, it becomes difficult to correct for any  density-dependent shifts in the collective responses~\cite{Zhu2016,Deb2020,Agarwal2024}.

\begin{figure}[htbp]
\centering
\includegraphics[width=\linewidth]{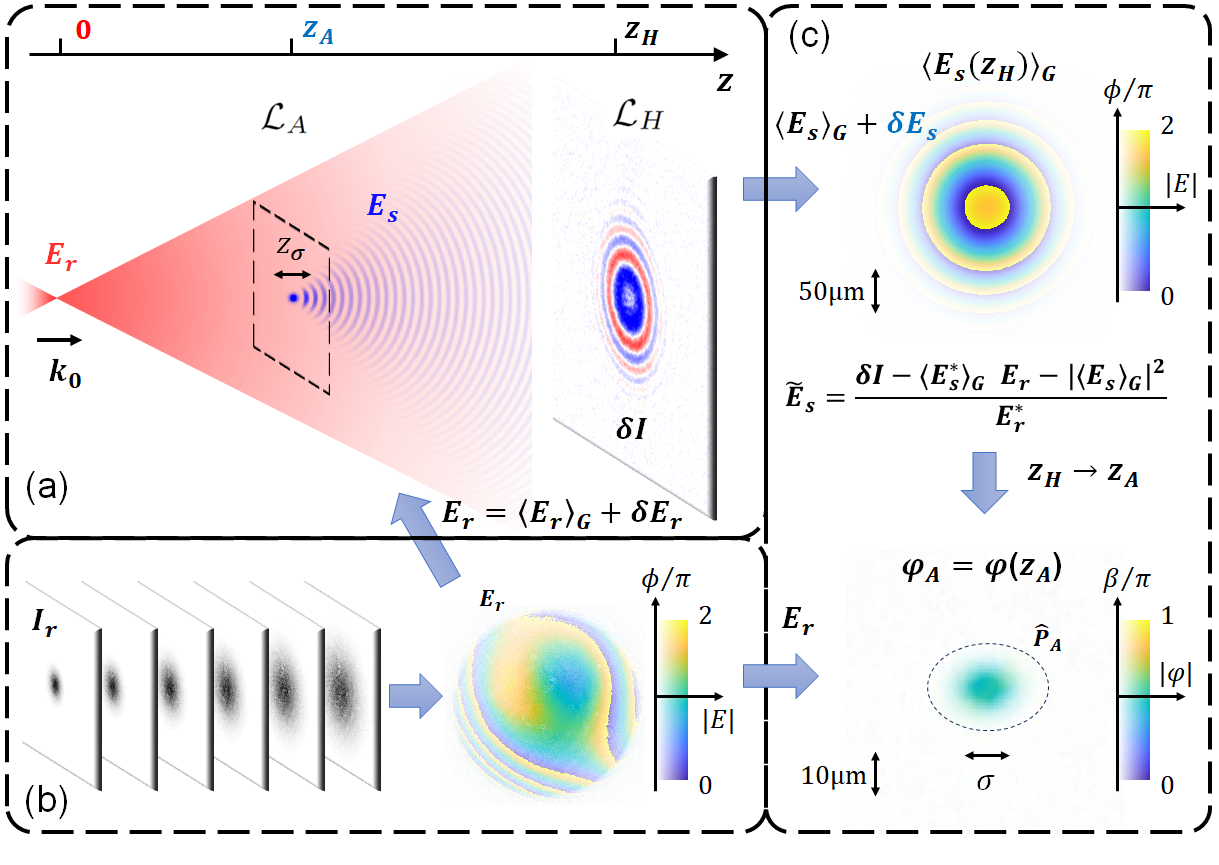}
\caption{Complex spectroscopic imaging with Gaussian-decomposition-assisted inline holography. (a): The basic setup. (b): Pre-characterization of the reference field $E_r=\sqrt{I_r}e^{i\phi_r}$ through multi-plane intensity $I_r=|E_r|^2$ measurements~\cite{Wang2022b}. The low-spatial-frequency part of the forward scattering $E_s=E_r'-E_r$ is decomposed into multiple Gaussian profiles according to Eq.~(\ref{eq:EG}). The Gaussian propagation supports efficient minimization of $\mathcal{L}_H$ associated with the $\delta I=I'-I$ data in the camera plane $z=z_H$, and $\mathcal{L}_A$ associated with the sample knowledge around $z=z_A$, according to Eq.~(\ref{eq:loss}). (c): The full $E_s$, approximated by $\tilde E_s$ in Eq.~(\ref{eq:halfstep}), is propagated together with $E_r$ to $z=z_A$ to retrieve the complex phase shift $\varphi(z_A)$ as the complex-valued imaging data. The aperture operator $\hat P_A$ encloses the sample with size $\sigma$ at $z=z_A$. The scale bars are according to typical experimental condition in this work. The (colored ) black symbols refer to quantities that are (not) precisely known at the represented stages of measurements. In (a) $z_{\sigma}=\pi \sigma^2/\lambda$.
}
\label{Fig1}
\end{figure}

In this work, we develop a systematic procedure for complex-valued spectroscopic imaging of laser-cooled atoms. The technique is based on inline holography~\cite{Gabor1972,Latychevskaia2019,Sobol2014, Wang2022b,Zhao2022b}. To fix the optical phase ambiguity~\cite{Gabor1946} by exploiting the prior knowledge about the atoms in a numerically efficient manner, we introduce a Gaussian-decomposition~\cite{Heller1975,Harvey2015, Ashcraft2020} method to digital holographic microscopy. For sparse array of small atomic samples, our method retrieves the transmitted $E_r'$ (and $E_r$) from single-shot hologram data.  By analyzing the phase angle $\beta$ near an isolated optical transition---which is highly resilient to power-broadening effects~\cite{Zhao2022b}---the centers of each sample can be precisely localized in 3D with super-resolution close to the photon-shot-noise limit (Cramér-Rao bound~\cite{Braunstein1996}). Absorption (${ \rm OD}$) and phase-shift ($\phi$) images are numerically reconstructed to form complex images $\varphi_A = \phi + i{\rm OD}/2$, achieving diffraction-limited spatial resolution and photon-shot-noise-limited sensitivity. Experimentally, we simultaneously reconstruct complex-valued spectroscopic images of sparsely distributed $^{87}$Rb samples on the D2 line, with micrometer-level 3D spatial resolution, from single-shot holograms recorded within tens of microseconds. By retrieving phase angles $\beta={\rm arg}(\varphi_A)$, we achieve a hundred-kHz-level resolution for measuring the transition frequency, using only hundreds of atoms, in a 3D-resolved manner. This is despite significant shot-to-shot atom-number variations as being unveiled by the modulus $|\varphi_A|$ readouts. The observed probe-induced Doppler shift can be corrected or mitigated~\cite{Zhao2022b} in future work.

In what follows, we first review the basic principles of the quantitative holographic imaging~\cite{Cuche1999, Marquet2005a, Nguyen2022} for atomic spectroscopy~\cite{Wang2022b}. We then introduce our new method where Gaussian decomposition~\cite{Harvey2015, Ashcraft2020,Heller1975} is applied into holography, to facilitate the phase retrieval and 3D localization of atomic samples. Finally, we summarize our motivation on holographic imaging for cold atoms.

\subsection{Complex-valued imaging with inline holography}\label{sec:cimg}
We consider the inline holography setup~\cite{Gabor1946,Latychevskaia2019} illustrated in Fig.~\ref{Fig1}a where a nearly spherical atomic sample at $z=z_A$, with a characteristic width $\sigma$, is resonantly excited by a spherical probe beam $E_r$ with optical frequency $\omega$, wavelength $\lambda$ and wavenumber $k_0=2\pi/\lambda$. The attenuation and phase shift to $E_r$ is encode in $E_s$, the coherent forward scattering from the sample. We therefore rewrite $E_r'=E_r+E_s$. Defining~\cite{foot:abre}
\begin{equation}
\varphi(z)=-i{\rm log}(1+E_s(z)/E_r(z)), \label{eq:varphi}
\end{equation}
then, according to Beer-Lambert law, we have the complex phase shift by the sample~\cite{Wang2022b}
\begin{equation}
\begin{aligned}
 \varphi_A(x,y)&=\varphi(x,y,z_A),\\
               &=\phi(x,y)+i{\rm OD}(x,y)/2,\\
  &\approx \frac{1}{2}k_0 \rho_c(x,y)\alpha(\omega).
  \end{aligned}\label{eq:varphiR}
\end{equation}
Here $\rho_c=\int \rho(z) {\rm d}z$ is the column density. The $\alpha(\omega)$ is the atomic polarizability. The third line in Eq.~(\ref{eq:varphiR}) becomes accurate for small ensemble of dilute atoms with $\rho\ll k_0^3$. 




The transmitted $E_r, E_r'$ propagate further for a distance $z_H-z_A$ where the intensity is recorded by a digital camera. The intensities in presence and absence of the atomic sample are modeled as
\begin{equation}
\begin{aligned}
I'&=\left|E_r(z_H)+E_s(z_H)\right|^2,\\
I&=\left|E_r(z_H)\right|^2,
\end{aligned}\label{eq:dI} 
\end{equation}
respectively. The reduced hologram is defined as
\begin{equation}
\delta I=I'-I.\label{eq:dI1}
\end{equation}
Assuming the probe profile $E_r$ as known, the goal of holographic imaging is to infer through the $I',I$ measurements the axial central location $z_A$ and to retrieve the 2D complex phase shift image $\varphi_A(x,y)$ (Eq.~(\ref{eq:varphiR})) there. 



\subsection{Challenges}\label{sec:challenge}
The quantitative holographic imaging technique~\cite{Cuche1999, Marquet2005a, Nguyen2022} faces  
at least three categories of challenges, which are all amplified in cold atom setups~\cite{Kadlecek2001,Turner2005, Sobol2014, Smits2020, Altuntas2021,Wang2022b,Zhao2022b}:

\begin{enumerate}
\item [C1:] To retrieve $E_s$ from the $I',I$ data with Eq.~(\ref{eq:dI}) is an ill-defined problem, requiring additional constraints to fix the $E_s$ solution not to be perturbed by its twin~\cite{Gabor1946}, 
\begin{equation}
E_{s,{\rm twin}}=\frac{E_r}{E_r^*} E_s^* \label{eq:twin}
\end{equation} 
at $z=z_H$, which focuses {\it approximately} at
\begin{equation}
z_{\rm twin}=\frac{z_A z_H}{2z_A-z_H}.\label{eq:zTwin}
\end{equation}
Due to the fragile nature of cold atomic samples and typical complexity of the setup that produce them, traditional solutions to the twin-image problem associated with phase-shifting~\cite{Yamaguchi1997} and multiple phase-diversity measurements~\cite{Paxman1992,Ou2013, Perry2021a} can not be easily implemented. While off-axes holography~\cite{Leith1962,Kadlecek2001,Smits2020} is powerful for resolving the optical phase, the technique necessarily sacrifices some of the imaging bandwidth while adding more complexity to the setup. Finally, while structured illumination~\cite{Osten2013} may help solving the phase problem, the associated $E_r$ structure could complicate the light-atom interaction through Doppler shifts and transition saturation~\cite{Brown2013a}. 

Without resorting to additional measurements, the constraints available for cold atom holography typically only include those associated with finite spatial support~\cite{Kadlecek2001,Sobol2014,Smits2020,Wang2022b} and optical response characteristics~\cite{Turner2005,Altuntas2021}.  Previous applications of these prior knowledge, while leading to notable successes, may still not guarantee an accuracy for precision spectroscopy in presence of imaging noises.

\item [C2:] The sample plane location $z_A$ is not known {\it a priori} for the numerical back-propagation of the reconstructed $E_r,E_s$~\cite{Smits2020,Zhao2022b}. Retrieving ${\rm OD}$ and $\phi$ with the Eqs.~(\ref{eq:varphi})(\ref{eq:varphiR}) relation in an out-of-focus plane mixes the absorption with phase shift~\cite{Zuo2020,foot:TIE}, leading to erroneous phase angle $\beta$.

Cold atoms in a dilute enough gas typically follow a normal density distribution. The lack of distinct spatial features invalidates a class of traditional methods for sample-plane localization in digital holography~\cite{ ilhan2013autofocusing, Brady2009,zhang2017edge,Fan2017,Wu2018a}. While the sample plane can be precisely located by observing the atomic shot noise~\cite{Altuntas2021}, the method is slow since many images are required for the statistical analysis.

\item [C3:] While the $E_r$ wavefront is assumed known, any deviation of the actual $E_r$ induces distortion to the retrieved $E_s$, leading to erroneous $\varphi_A$.  In particular, the speckle noises~\cite{Redding2012} in $E_r$ can hardly be avoided in atomic spectroscopy where the laser probe is necessarily monochromatic~\cite{Sobol2014,Wang2022b}. 
\end{enumerate}


\subsection{This work}
We develop a method for complex-valued spectroscopic imaging of sparsely distributed atomic samples in 3D.  As outlined in Fig.~\ref{Fig1}, the method is based on inline holography~\cite{Gabor1972}. The phase ambiguity associated with the twin image (Eq.~(\ref{eq:twin})) is addressed by systematically exploiting our prior knowledge about the measurement. Apart from the hologram data (Eq.~(\ref{eq:dI})), the pre-characterized $E_r$ wavefront~\cite{Wang2022b}, our prior knowledge is mostly about the bulk optical properties of the samples in focus, {\it i.e.}, aspects of the complex phase shift profile $\varphi_A(x,y)$ at $z=z_A$. The key insight is that when the probe beam $E_r$ seen by the sample is locally smooth, then these properties can be efficiently parametrized by decomposing $E_s$ into a few Gaussian beams (see Eq.~(\ref{eq:G}))~\cite{Heller1975,    Harvey2015, Ashcraft2020},
\begin{equation}
E_G ({\bf r};\{c_j,{\bf P}_j\})=\sum_{j=1}^{n_A} c_j G_j({\bf r};{\bf P}_j).\label{eq:EG}
\end{equation}
We denote the optimal decomposition as $\langle E_s\rangle_G=E_G ({\bf r};\{c_j,{\bf P}_j\}_{\rm opt})$. The Gaussian propagation facilitates simultaneous minimization of cost functions,  $\mathcal{L}_H$ at $z=z_H$ (Eq.~(\ref{eq:lossH})) associated with the $I',I$-data, and $\mathcal{L}_A$ at $z=z_A$ (Eq.~(\ref{eq:lossC})) associated with the expected $\varphi_A$. 

Our method works especially well in a ``defocused twin regime'' (Appendix~\ref{sec:dftwin}) for small atomic samples, with $z_{\sigma}\ll z_A, z_H-z_A$, so that there are sufficient interference fringes in the $I'-$data (Eq.~(\ref{eq:dI1})). Here  $z_{\sigma}=\pi\sigma^2/\lambda$ is the characteristic Rayleigh length of the sample. Key advantages of our method are summarized as follows:

\begin{enumerate}
    \item [A1:] Our method generalizes traditional single-shot twin-removal strategies such as those associated with finite spatial supports~\cite{Koren1993,Isikman2011,Sobol2014} and spectral phase angles~\cite{Turner2005,Latychevskaia2007}.  Here, bulk properties of the samples such as the approximate 3D locations, shapes, and spatial-(in)dependent spectral responses, can all be encoded into the $\varphi_A$-image-expectations to constrain the $\{c_j, {\bf P}_j\}$ parameters. The enhanced application of prior knowledge improves the accuracy of the optical phase determination in presence of imaging noises.

    \item [A2:] With the phase angle $\beta={\rm arg}(\varphi_A)$ constrained by knowledge of light-atom interaction~\cite{Turner2005,Altuntas2021,Zhao2022b}, the atomic sample plane location $z_A$ can be conveniently inferred during Gaussian decomposition. For nearly spherical samples with $z_{\rm \sigma}\ll z_A, z_H-z_A$, the typical axial resolution is~\cite{Zhao2022b} (Appendix~\ref{sec:Fisher})    
    \begin{equation}
    \Delta z_A\approx z_{\sigma}/\sqrt{N_s},\label{eq:dz}
    \end{equation}
    decided by the Rayleigh length $z_{\sigma}=\pi\sigma^2/\lambda$ associated with the sample size $\sigma$ (Fig.~\ref{Fig1}a). This resolution can be enhanced by introducing finer sample features, or simply by repeating the in situ measurements with standard samples to effectively increase the total number of coherently scattered photons  
  $N_s=\sum_{x,y} |E_s|^2$ received by the camera.
    
    
    \item [A3:]  With $z_{\sigma}\ll z_A, z_H-z_A$, diffraction-limited $E_r$ and $E_s$  can be retrieved from single-shot $I',I$ data, using the pre-characterized $E_r$ knowledge through multi-plane intensity measurements~\cite{Wang2022b}. Impacts of mild speckle noises and  moderate aberrations can be corrected, for achieving complex $\varphi_A(x,y)$
    imaging with diffraction-limited resolution and photon-shot-noise limited sensitivity. The phase-angle resolution is given by~\cite{Wang2022b} (Appendix~\ref{sec:Fisher})    
    \begin{equation}
    \Delta \beta(\mathcal{A})\approx 1/\sqrt{2 N_s(\mathcal{A})}.\label{eq:db}
    \end{equation}
Here $N_s(\mathcal{A})=\sum_{\mathcal{A}} |E_s|^2$ integrates the elastically scattered photons over area $\mathcal{A}$ of interest in the in-focus $\varphi_A(x,y)$ image.  
   \end{enumerate}



\subsection{Application to atomic spectroscopic imaging}\label{sec:spec}
With the complex $\varphi_A=|\varphi_A|e^{i\beta}$ readouts as spectroscopy data, the previously mentioned challenges in cold atom spectroscopy—stemming from fluctuations in atom number and light-atom interaction strength—become significantly more manageable. When combined with 3D imaging capabilities, as well as the power-broadening resilience~\cite{Zhao2022b} to be discussed in the following, our complex spectroscopy technique substantially improves the potential for imaging and sensing on strong transitions with laser-cooled atomic arrays.

\subsubsection{Normalization-free phase-angle spectroscopy}\label{sec:normf}
In traditional spectroscopy, to retrieve single-atom spectroscopic properties from collective readouts, normalization of atom number and light-atom interaction strength is typically required. This normalization is usually achieved by following the spectroscopic measurements with an additional measurement, where the probe laser operates under standard conditions~\cite{Sansonetti2011,Li2020}. Except for setups with a known atom number and fixed interaction strength~\cite{Marti2018,stamperkurn}, this normalization step is generally necessary, which increases measurement noise, as will be discussed shortly. Additionally, mitigating measurement-induced back-action becomes crucial for samples subjected to multiple measurements.

Here, according to Eq.~(\ref{eq:varphiR}), for small and dilute samples in the Beer-Lambert regime the probe transmission is determined by the complex phase shift $\varphi_A(x,y) \propto \rho_c(x,y)\alpha$. The phase angle $\beta$ can be obtained from the OD and $\phi$ data derived from single-shots without additional measurements. The value of $\beta$ reflects the phase angle of the atomic polarizability $\alpha$, and is insensitive to the number of atoms, nor the detailed atomic population distribution in the ground states which determines the light-atom interaction strength (which is prone to optical pumping~\cite{Happer1972}, Appendix~\ref{sec:Fig2detail}). The atom-number and interaction-strength variations can be inferred from the modulus $|\varphi_A(x,y)|$. Beyond the single-atom picture, density-dependent shifts~\cite{Zhu2016}  can be modeled according to the OD and $\phi$ image data.

\subsubsection{Shot-noise-limited precision}\label{sec:spec1}
As a heterodyne measurement, holographic imaging inherently suffers from a 50\% loss of Fisher information~\cite{Kadlecek2001, Sobol2014} (see Appendix~\ref{sec:Fisher}) in the transmitted probe. Therefore, absorption~\cite{Hung2011} and phase-contrast~\cite{Higbie2005} techniques are fundamentally more effective for imaging absorptive and phase samples, respectively. However, our complex spectroscopy method directly extracts spectroscopic information from the phase angle $\beta$, thereby circumventing the measurement noise associated with normalizing atom number and interaction strength. If this normalization measurement is replaced by an additional, independent holographic measurement, the Fisher information is effectively restored. Therefore, we expect complex spectroscopic imaging to achieve photon-shot-noise-limited performance comparable to that of absorption techniques for absorptive samples and phase-contrast techniques for phase samples. In practice, the phase angle is often unknown and signal normalization is inherently imperfect. Therefore, the holographic complex-valued spectroscopy, if works, is likely to outperform both absorption and phase-contrast measurements.




\subsubsection{Power-broadening resilience}\label{sec:cPhase2level}

\begin{figure}[htbp]
\centering
\includegraphics[width=\linewidth]{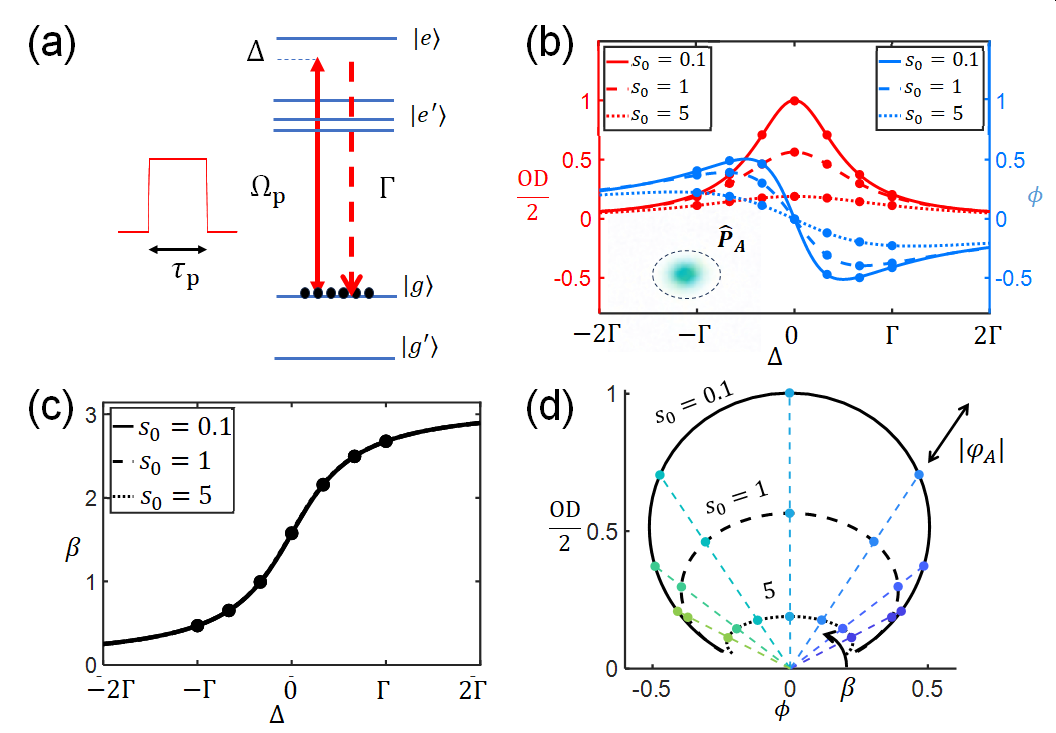}
\caption{Power-broadening resilience in normalization-free phase-angle atomic spectroscopy. (a): Energy diagram for atomic spectroscopy on $|g\rangle-|e\rangle$ transition, corresponding to the $F=2 - F'=3$ hyperfine D2 transition of $^{87}$Rb in this work. (b): Simulated average ${\rm OD}=2{\rm Im}(\varphi_A)$ and $\phi={\rm Re}(\varphi_A)$ with $\tau_{\rm p}=10~\mu$s square-pulse excitation, similar to the experimental $\tau_{\rm p}=20~\mu$s pulse in Sec.~\ref{sec:exp}. Practically, one averages the reconstructed $\varphi_A$ as those in Fig.~\ref{Fig1}b within the aperture $\hat P_A$ (or any specific regime-of-interest), as suggested by the inset, to obtain $\varphi_A$ (denoted as $\bar\varphi_A$ in Fig.~\ref{Fig4}). The same simulated data generates phase-angle spectroscopy $\beta$ in (c) and (d) to show strong power-broadening resilience. The $s_0-$dependent deviations below $10^{-2}$ are invisible in (c) (Appendix~\ref{sec:Fig2detail}~\cite{foot:future}). In addition, the magnitude $|\varphi_A|$ depends on atomic density $\rho_c$ and light-atom interaction strength $\eta^2$ (Appendix~\ref{sec:Fig2detail}), which practically often fluctuate as suggested by the double-sided arrow in (d) (Also see Fig.~\ref{Fig4})). 
}
\label{FigSat}
\end{figure}

Spectroscopic measurements depend on accurate modeling of light-atom interactions to extract valuable information. Although the single-atom linear response, denoted by $\alpha$ in Eq.~(\ref{eq:varphiR}), can be modeled perturbatively~\cite{Meppelink2010}, the precision of this estimation is often affected by optical saturation~\cite{Brown2013a} and multi-scattering~\cite{Zhu2016} effects. A closer examination of the transient atomic response indicates that, even with a relatively strong probe—whether to suppress multi-scattering effects~\cite{Reinaudi2007,Chomaz2012} or to enhance the signal—the complex phase can still often be approximated as $\varphi_A \approx \frac{1}{2}k_0\rho_c\alpha$ (Eq.~\eqref{eq:varphiR}), with (Appendix~\ref{sec:OBE})
\begin{equation}
    \alpha = \xi(s_0,\Delta) \frac{3}{8\pi^2} \lambda^3 \frac{-\Gamma/2} {\Delta + i\Gamma/2}.\label{eq:varphi2level}
\end{equation}
Here, $s_0$ is the probe saturation parameter, and $\Delta = \omega - \omega_{eg}$ is the detuning of the probe from the atomic resonance frequency $\omega_{eg}$ under consideration (Fig.~\ref{FigSat}a). The linewidth is denoted by $\Gamma$. The real number $\xi(s_0,\Delta)$ is a saturation factor related to the atomic internal state dynamics. For instance, for a sufficiently long probe on a closed transition, allowing the atomic response to reach a steady state, we obtain the standard $\xi(s_0,\Delta)\approx \eta^2/(1+s)$ associated with coherent dipolar excitation. Here, $\eta^2 \leq 1$ is determined by the probe polarization and the optical-pumping induced atomic population redistribution (Appendix~\ref{sec:Fig2detail}), and $s = s_0 \Gamma^2/(4\Delta^2+\Gamma^2)$~\cite{Steck2003}.


The $\xi(s_0,\Delta)$ factor leads to the power-broadening of absorption ${\rm OD}(\Delta)$ and phase shift $\phi(\Delta)$ lineshapes. Similar broadening is anticipated in fluorescence spectroscopy~\cite{WolfBook}. 
Due to the sensitivity of $\xi(s_0, \Delta)$ in Eq.~(\ref{eq:varphi2level}) to both the local laser intensity and the characteristics of the probe waveform, to accurately model this factor is difficult practically. A common approach in precision spectroscopy is to perform multiple frequency scans and then extrapolate the results to the weak probe limit~\cite{Brown2013a}. This slow process often limits the bandwidth of precise measurement and sensing.

On the other hand, from Eq.~(\ref{eq:varphi2level}) it is evident that as long as $\xi(s_0,\Delta)$ remains real, the phase angle $\beta = \arg[\alpha]$ is not influenced by power broadening even when $s \gg 1$. In Appendix~\ref{sec:OBE} we clarified~\cite{Zhao2022b} that in absence of the manageable Doppler shifts (Appendix~\ref{sec:Doppler}),
the $\xi(s_0,\Delta)$ remains real for spectrally isolated $g-e$ transitions, whether open or closed, as long as virtual mixing to distant levels ($|e'\rangle$, $|g'\rangle$ in Fig.~\ref{FigSat}a) is negligible, and for smooth, long probe pulses with duration $\tau_{\rm p}\gg 1/\Gamma$, such that transient effects can be averaged out. 

In Fig.~\ref{FigSat} we verify the robustness of the phase angle $\beta$ by numerically evaluating the atomic polarizability $\alpha$ for $^{87}$Rb atoms, resonantly excited on the $F=2 \rightarrow F'=3$ hyperfine transition of the D2 line, which will also be examined experimentally. The probe pulse $E_r$ is characterized by a duration $\tau_{\rm p} \gg 1/\Gamma$ and a Rabi frequency $\Omega_{\rm p} = \sqrt{s_0/2}\Gamma$.  As shown in Fig.~\ref{FigSat}(c,d), the phase angle $\beta$ remains insensitive to saturation effects as $s_0$ increases, in stark contrast to the significant power broadening observed in the absorption ($\mathrm{OD}$) and phase shift ($\phi$) shown in Fig.~\ref{FigSat}b. Further details of the simulation can be found in Appendix~\ref{sec:OBE}~\cite{foot:future}.

%


\subsubsection{3D imaging and sensing}\label{sec:multiI}

The scenario depicted in Fig.~\ref{Fig1}a, which involves a single sample, can be generalized to multiple samples. As detailed in Appendix~\ref{sec:multipleA}, if the samples are sufficiently sparse such that the forward scatterings, $E_{s,A}$, from each sample $A$ are distinguishable, then, given the complex fields $E_r$ and $E_s = \sum_A E_{s,A}$, the holographic technique can reconstruct the individual optical responses of multiple samples distributed in 3D. Furthermore, when combined with faithful $\beta$-readouts that are resilient to power broadening, the normalization-free measurement enables transition-frequency resolution for spectroscopic sensing of 3D potentials, within a single shot, as will be demonstrated in Sec.~\ref{sec:exp}. 

We note that diffraction-limited imaging with 3D resolution can be highly useful in ultracold atomic physics research~\cite{Nelson2007, Eliasson2020, Legrand2024}. On the frequency resolution side, accurate spectroscopic imaging has been demonstrated on narrow-line transitions~\cite{Marti2018, Diego2023}, or with single-atom-based scanning probe techniques~\cite{stamperkurn}. In comparison to these previous approaches, our ensemble-based strong-transition measurements substantially expand the time-frequency window for quantum sensing with cold atoms, potentially over large volume-of-view~\cite{Sobol2014a}.

\subsection{Major limitations in this work}\label{sec:limitation}
Practically, our holographic imaging method~\cite{Wang2022b} can be conveniently adapted to standard imaging setups, simply by translating the camera away from the atomic imaging plane for recording the inline interference patterns (Fig.~\ref{Fig1}a). Numerically, a key step is based on matching $E_r, E_s$ with free-propagating Gaussian beams through the nonlinear minimization of Eq.~(\ref{eq:loss}). The method helps to overcome several major obstacles in quantitative holography, but also introduces substantial limitations. In the following we discuss major limitations and possible improvements. 

First, the nonlinear optimization is quite slow. As we find in this work, while the analytical form of the Gaussian beam (Eq.~(\ref{eq:G}))~\cite{Harvey2015, Ashcraft2020,Heller1975} helps to avoid the resource-demanding, numerical propagation of full wavefronts~\cite{Sobol2014,Wang2022b,Smits2020,Altuntas2021}, to resolve single hologram still requires about 2 minutes even with a A100 GPU card (NVIDIA). Notably, for repeated measurements such as during a probe-frequency scan~\cite{Sansonetti2011,Li2020},  the processing time can be substantially reduced, down to tens of seconds, by seeding the optimization process with nearly optimal initial $\{c_j,{\bf P}_j\}$-parameters. With more advanced software and hardware supports, we expect further improved imaging processing speed in future work.  

Second, as detailed in Appendix~\ref{sec:dftwin}, diffraction-limited $\varphi(z_A)-$image is most easily achieved for small samples satisfying $z_{\sigma}\ll z_A,z_H-z_A$ (Fig.~\ref{Fig1}a) only, or, equivalently, with the sample size $\sigma\ll \sqrt{\lambda z_A},\sqrt{\lambda (z_H-z_A)}$. Beyond this ``defocused-twin regime'', accurate $\varphi_A-$retrieval relies on improving the decomposition of $E_s$ using more Gaussian profiles (larger $n_A$), {\it i.e.},  to capture fine details of $E_s$ so that the defocused twin-image noise (Eq.~(\ref{eq:halfstep})) around $z=z_A$ is more efficiently removed. While the principle of such improvement as outlined in Sec.~\ref{subsec:optimize} appears straightforward, its practicality in presence of imaging noise requires systematic study in future work. Nevertheless, we note that depending on the spatial resolution requirements, the $z_A$ and $z_H-z_A$ limited by the camera sensor size can typically be at a centimeter level. This in term supports sparse arrays with individual sample size of up to $100~\mu $m (Fig.~\ref{Fig6}), large enough for various applications. At moderate $z_{\sigma}$, the Gaussian-decomposition method can also be assisted by Gerchberg-Saxton-type iterations~\cite{Koren1993, Sobol2014, Latychevskaia2019,
Wang2022b} to improve the $E_s$ estimation.


Finally, the Gaussian-assisted holography method in this work assumes free-propagation between the sample plane at $z=z_A$ and camera plane at $z=z_H$. Practically, the atomic samples are often imaged to the camera plane through a lens array~\cite{Hung2011, Kuhr2016,Altuntas2021}. In this experimental work, the moderate, aplanatic aberrations in $E_s$ are corrected by a pre-characterized point-spread-function near the atomic sample location (Appendix~\ref{sec:aberrApp}). Obviously, the method becomes insufficient in presence of strong astigmatism during, {\it e.g.}, wide-field imaging~\cite{Zheng2013,Zheng2013a,Park2021}. In future work and with sufficient knowledge of the imaging system, full suppression of strong, astigmatic aberrations can be achieved by exploiting the Gaussian propagation across the lens array numerically~\cite{Harvey2015, Ashcraft2020}. This may potentially enable diffraction-limited 3D complex-valued spectroscopic imaging across unprecedented imaging volumes, only limited by the pixel size of the camera~\cite{Sobol2014a}. 



\subsection{Structure of the rest of the paper}
In the following the main text is structured into three sections. With technical details presented in Appendix~\ref{sec:scheme}, we first outline the principles of the Gaussian-decomposition-assisted holographic imaging method in Sec.~\ref{sec:principle}. Next, in Sec.~\ref{sec:exp} we present experimental demonstration of diffraction and shot-noise limited complex-valued imaging. We summarize this work in Sec.~\ref{sec:conclusion}. 

\section{Principles}\label{sec:principle}


The schematic setup for our Gaussian-decomposition-assisted complex-valued spectroscopic imaging method is shown in Fig.~\ref{Fig1}(a-c). The setup in Fig.~\ref{Fig1}(a) has already been described in Sec.~\ref{sec:cimg}. We consider the $(I', I)$ measurement to be a single-shot process. Since no atom-number or interaction-strength normalization is required afterward, the $I'$ exposure can be strong and prolonged, which may significantly alter both the atomic velocity and the internal-state population distribution. The $I$ exposure is applied immediately after the atomic sample is dispersed away from the $E_r$ illumination.

We assume the holographic imaging setup to be effectively lensless~\cite{Sobol2014} so that the accurate propagation of wavefronts between $z_A$ and $z_H$ is modeled by Eq.~(\ref{eq:asm}), $E(z')= \hat U(z'-z)E(z)$, through angular spectrum method numerically. On the other hand, for repetitive short-distance propagation during numerical optimization, the Gaussian-decomposed  $E_r, E_s$ wavefronts are sampled according to the analytical Eq.~(\ref{eq:G}) under paraaxial approximation (also see Appendix~\ref{sec:aberrApp}.).

The reconstruction of $\varphi_A(x,y)$ 
at $z=z_A$ is divided into Sec.~\ref{sec:Er},\ref{subsec:optimize},\ref{sec:step3} steps.

\subsection{Obtaining $E_r$}\label{sec:Er}
A prerequisite to recover $E_s$ through the holographic measurement with Eq.~\eqref{eq:dI} is to characterize the $E_r$ wavefront precisely. The steps for the $E_r$ measurements are illustrated in Fig.~\ref{Fig1}(a,b), detailed in ref.~\cite{Wang2022b}. Briefly, using the single-shot $I$-data, the probe wavefront $E_r=\sqrt{I}e^{i\phi_r}$ is obtained by intensity-matching $I$ to $I_0=|E_r^{(0)}(z_H)|^2$. Here $\phi_r={\rm arg}(E_r^{(0)})$. $E_r^{(0)}$ is the pre-characterized probe wavefront, obtained with multi-plane intensity measurements of $E_r$, followed by Gerchberg-Saxton iteration~\cite{Yang1994} (Fig.~\ref{Fig1}b). Notice the $E_r$ wavefront measurement is designed to share the same optical path with the holographic measurement. 

\subsection{Minimal Gaussian-decomposition of $E_s$}
\label{subsec:optimize}

We parametrize $E_s$ with $n_A$ Gaussian beams according to Eq.~(\ref{eq:EG}), and minimize the cost function 
\begin{equation}
  \mathcal{L}_{\rm tot}(z_A,\{c_j,{\bf P}_j\}) =\mathcal{L}_H+\mathcal{L}_A,\label{eq:loss}
  \end{equation}
with
\begin{equation}
\mathcal{L}_H(\{c_j,{\bf P}_j\})=\sum_{x,y}\left|\delta I-E_{r}^* E_G-E_{r} E_G^* - |E_G|^2\right|^2,  \label{eq:lossH}
  \end{equation}
and
\begin{equation}
\mathcal{L}_A(z_A,\{c_j,{\bf P}_j\})= \sum_{\mu }m_{\mu} C_{\mu}\left(\varphi_G(z_A,\{c_j,{\bf P}_j\})\right).\label{eq:lossC}
\end{equation}
Here $\mathcal{L}_A$ is formulated into a list of $C_{\mu}$-constraints (Appendix~\ref{sec:SC}) weighted by $m_{\mu}$, which can be adjusted during the optimization. The $\varphi_G(z_A;\{c_j,{\bf P}_j\}) =-i{\rm log}\left(1+E_G/\langle E_r \rangle_G \right)_{z=z_A}$ evaluates the complex-valued phase shift by Eq.~(\ref{eq:varphi}), with $E_r$ replaced by the known Gaussian envelope $E_r\approx \langle E_r\rangle_G$.
Simultaneous evaluation of Eq.~(\ref{eq:lossH}) at $z=z_H$ and (\ref{eq:lossC}) with varying $z_A$ is assisted by the analytical Gaussian propagation (Eq.~(\ref{eq:G})). 

We assume the envelope $\langle E_s\rangle_G$ to be sparse in the Gaussian beam basis. To ensure a minimal $n_A$ for the $E_s$ decomposition, we start with $n_A=1$ in Eq.~(\ref{eq:EG}) for each atomic sample and progressively increase $n_A$ by splitting one of the optimal Gaussian beams for additional optimization. During the process, the quality of the decomposition is checked by analyzing the difference between $\langle E_s(z_A)\rangle_G$ with the $\tilde E_s(z_A)$ approximation, 
\begin{equation}
\begin{aligned}
     \tilde E_s(z_A) &=\hat U (z_A-z_H) \tilde E_s(z_H),~{\rm with}\\
         \tilde E_s(z_H)&=\left(\frac{\delta I- E_{r} \langle E_s \rangle_G^* -|\langle E_s \rangle_G|^2}{E_r^*}\right)_{z=z_H}
    \end{aligned}\label{eq:halfstep}
\end{equation}
as
\begin{equation}
\delta E_G(z_A)= \tilde E_s(z_A)-\langle E_s(z_A)\rangle_G.\label{eq:dE}
\end{equation} 
Here the Gaussian approximation $\langle E_s(z_A)\rangle_G$ only contains low-spatial-frequency part of $E_s(z_A)$. Instead,  $\tilde E_s(z_A)$ has the full imaging bandwidth, except that with the twin (Eq.~(\ref{eq:twin})) and a typically weaker ``dc'' $|E_s|^2/E_r^*$ approximately removed, $\tilde E_s(z_A)$ is subjected to disturbance by high-frequency residual of $E_{s,{\rm twin}}$  not captured by the Gaussian approximation. In principle, the optimization should continue until sufficient $n_A$ Gaussians ensures fine enough decomposition, so that the $\delta E_G(z_A)$ in Eq.~(\ref{eq:dE}) becomes featureless, {\it i.e.}, the two approximations agree at low-spatial-frequency of interest. In this work, the decomposition process is terminated emphatically (Sec.~\ref{sec:exp}) at $n_A=3$ for small spherical samples. (See Appendix~\ref{sec:large} for larger $n_A$ examples). We expect
more systematic study in future work on the refinement of $\langle E_s\rangle_G$ decomposition, particularly for large samples.



Finally, as already being suggested in Sec.~\ref{sec:multiI}, the above analysis is easily generated to multiple samples, indexed by $A$, which are  centered at ${\bf r}_A=(x_A,y_A,z_A)$ to be sparse enough so that their coherent forward scatterings are substantially different (Appendix~\ref{sec:multipleA}). We refer the coherent forward scattering by each sample as $E_{s,A}$ so that $E_s=\sum_A E_{s,A}$.



\subsection{Retrieving diffraction-limited $\varphi_A$}\label{sec:step3}

When the constrained Gaussian-decomposition to $E_s$ is successful, an estimation of the Gaussian envelop $\langle E_s\rangle_G$ is obtained. Depending on the prior knowledge of the phase angle $\beta$, the sample plane location $z=z_A$ is determined with an accuracy close to the photon-shot-noise limit (Eq.~(\ref{eq:dz})) (Appendix~\ref{sec:Fisher}). We therefore obtain the low-resolution envelop of the complex-valued phase-shift,
\begin{equation}
    \langle \varphi_A \rangle_G=-i{\rm log}\left(1+\langle E_{s,A}\rangle_G/\langle E_r \rangle_G \right)_{z=z_A}.\label{eq:vphil}
\end{equation}

 
When $z_{\sigma}\ll z_A,z_H-z_A$, the low-resolution reconstruction of $\langle \varphi_A\rangle_G$ can be refined into diffraction-limited $\varphi_A$ image according to Eq.~(\ref{eq:varphi}), simply by replacing the $E_s$ there with
 \begin{equation}
E_{s,A}(z_A)\approx \hat P_A\left(\tilde E_{s}(z_A)\right),  \label{eq:EsApp}
\end{equation}
where
\begin{equation}
  E_r(z_A)= \hat U(z_A-z_H) E_r(z_H).\label{eq:Ercorr}
\end{equation}
The $z_A$ localization can be accordingly refined. The aperture operator $\hat P_A$ encloses a minimal area at $z=z_A$, according to our prior knowledge about the sample location (Fig.~\ref{Fig1}c), to select the $E_{s,A}$ wavefront. 

As to be explained in Appendix~\ref{sec:dftwin},  
$\tilde E_s(z_A)$ contains all $E_s$ features within the full imaging bandwidth, except that it is subjected to disturbance by the out-of-focus, high-frequency-part of $E_{s,{\rm twin}}$ not captured by the Gaussian approximation. By operating the Fig.~\ref{Fig1}a  setup in the ``defocused-twin regime'' with $z_{\sigma}\ll z_A,z_H-z_A$, the residuals spread out sufficiently not to affect the $\varphi_A$ retrieval. In addition, Eq.~(\ref{eq:EsApp}) assumes that the multiple samples are only weakly displaced along $z$, as in this experimental work, with coherent forward scattering not overlapping in the sample planes. To retrieve $\varphi_A$ for multiple samples with substantial $E_{s,A}$ overlap requires separation of each $E_{s,A}$ in a self-consistent manner, a scenario that will be discussed in Appendix~\ref{sec:multipleA}.

\section{Experimental demonstration}\label{sec:exp}
In this section, we demonstrate 3D complex-valued spectroscopic imaging of laser-cooled $^{87}$Rb samples on the D2 line. The energy diagram is given in Fig.~\ref{Fig2}a.  The D2 linewidth is $\Gamma=2\pi\times 6.1~$MHz~\cite{Steck2003}.

\subsection{Experimental setup}
\begin{figure}[htbp]
\centering
\includegraphics[width=\linewidth]{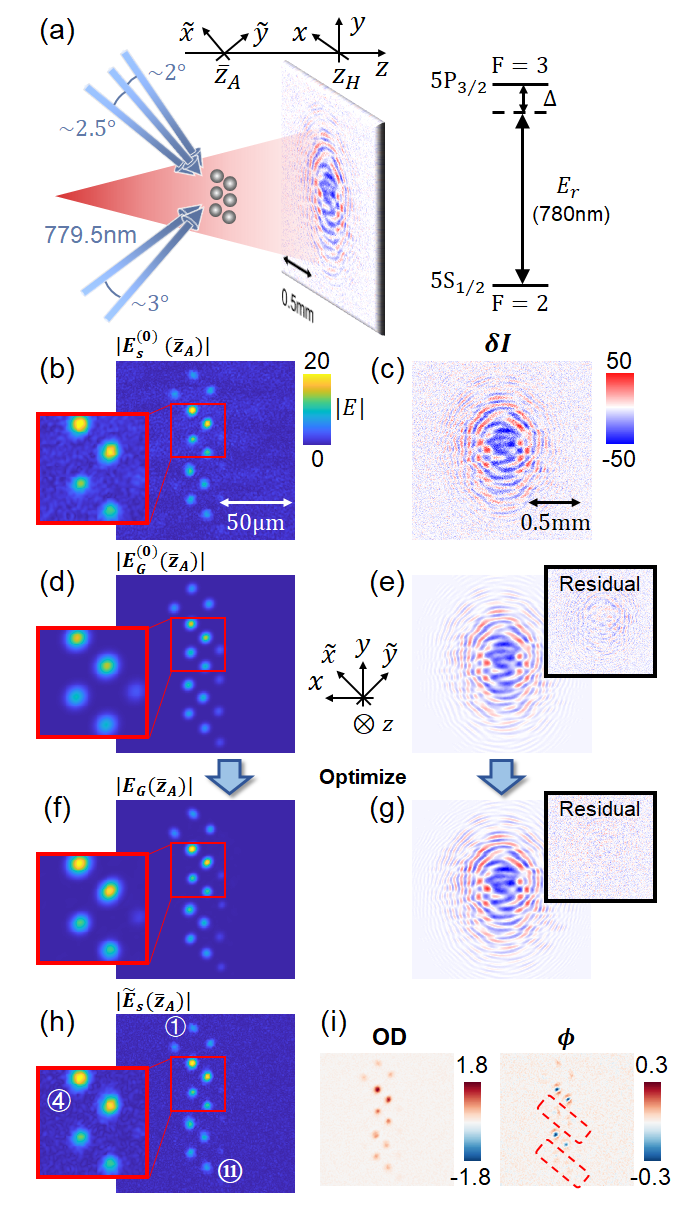}
\caption{
  (a): The experimental setup, typical hologram $\delta I$ at $\Delta= 0$~\cite{foot:LFNB}, and the level diagram. (b)-(c): Reduced hologram $\delta I$ (averaged over 9 exposures to enhance the display) and the modulus of  $E_{s}^{(0)}(\bar{z}_A)=\hat U(\bar{z}_A-z_H)\left(\frac{\delta I}{E_r^*}\right)$ in $\bar{z}_A$ plane. (d)-(e): Initial Gaussian profiles in the $z={\bar z}_A$ plane (plotted as $|E_G^{(0)}|$) and expected $\delta I_e$ in $z=z_H$. The sub-figure on (e) is the residual $\delta I_e-\delta I$. (f)-(g): Modulus of the optimal Gaussian profiles $E_G=\langle E_s(\bar{z}_A)\rangle_G$ and the corresponding hologram. (h): Modulus of $\tilde{E}_s(\bar{z}_A)$ with Gaussian-approximated twin-image removed. Eleven samples are labeled from left to right, top to bottom. (i): Reconstructed $\rm OD$, $\phi$ distributions. At $z=\bar{z}_A$, only the samples in dash-lined boxes are with the phase-shift $\phi$ rightly suppressed in-focus.
}
\label{Fig2}
\end{figure}

The schematic setup is illustrated in Fig.~\ref{Fig2}a. The probe beam is generated by an ${\rm NA}=0.3$ objective~\cite{Alt2002} to create a diffraction-limited focus at $z=0$ in vacuum. A sparse array of $^{87}$Rb atoms is created around $z_A\approx 1.5$~mm, as to be detailed in the following. During holographic imaging, the $E_r, E_s$ wavefronts are relayed by an aberration-compensated $D=50.8$~mm-diameter lens-array~\cite{LI2018}, not shown in the figure, to the camera (PICO Pixelfly-USB) sensor outside the vacuum. The imaging system has a reduced numerical aperture of ${\rm NA}=0.26$, limited by additional optics following the objective, and a lateral magnification of $M=3.5$. To perform inline holography, the camera is translated along $z$, defocused by $\Delta z_H^{(0)}=17.5$~mm, to record the magnified $I',I$ holograms. In the following and for the convenience of presentation, we describe the imaging process in-situ, as in Fig.~\ref{Fig2}a, where the camera plane is effectively at $z_H$ that is displaced from $z_A$ by $\Delta z_H^{(0)}/M^2\approx 1.4$~mm.  We refer readers to Sec.~\ref{sec:aberr} and Appendix~\ref{sec:aberrApp} where steps for aberration corrections are discussed. 

We construct a 3D optical lattice, as in Fig.~\ref{Fig2}a, to confine a sparse atomic sample array. The lattice geometry is adapted to the limited optical access in our setup. Two of the five lattice beams, propagating approximately along $\tilde {\bf e}_y=\frac{1}{\sqrt{2}}(-{\bf e}_x+{\bf e}_y)$ with a $3^{\circ}$ intersecting angle, creates 1D lattice along the $\tilde {\bf e}_x=\frac{1}{\sqrt{2}}({\bf e}_x+{\bf e}_y)$ direction. Three other beams propagate approximately along $\tilde {\bf e}_x$, with $2^{\circ}\sim 2.5^{\circ}$ opening angles to form a square lattice in the $\{\tilde {\bf e}_y,{\bf e}_z\}$ plane. The primitive lattice vectors are oriented along the $\frac{1}{\sqrt{2}}(\tilde {\bf e}_y\pm {\bf e}_z)$ directions. The lattice constants along $\tilde {\bf e}_x$ and in the $\tilde {\bf e}_y-{\bf e}_z$ plane are $15~\rm{\mu m}$, $18~\rm{\mu m}$ and $22~\rm{\mu m}$, respectively (Also see Fig.~\ref{Fig3}(d,e)).  A 20~MHz relative shift  between the two $\tilde {\bf e}_y-$beams and the three $\tilde {\bf e}_x-$beams removes the interference between the two lattice confinement dimensions.

We choose the lattice laser wavelength to be $\lambda_l=779.5~$nm, blue-detuned from the rubidium D2 line. With $\sim 100$~mW per beam, $10~$mK-level trapping potentials are formed for the ground-state rubidium atoms around the dark centers of lattices.  Following a gray molasses cooling stage~\cite{bruce2017}, 
$^{87}$Rb from a magneto-optical trap is loaded directly into the lattices, resulting in samples with a characteristic width of $\sigma\approx 2.5~\mu$m and typical atom number of $10^2\sim 10^3$ per site. Observing these samples along $z$, the transverse displacements between adjacent samples are about $10~\mu$m,  well resolvable with our coherent imaging resolution of $x_{\rm res}=\lambda/{\rm NA}=3~\mu$m~\cite{Sobol2014a}. Here $\lambda=780~$nm. On the other hand, the axial displacement of $10~\mu$m is below the diffraction-limited resolution of $z_{\rm res}=2\lambda/{\rm NA}^2=23~\mu$m. With $\rho=10^{12}\sim 10^{13}/{\rm cm}^3$ in density, we expect dipolar shifts to be quite negligible~\cite{Zhu2016} and our single atom picture assumed by Eq.~\eqref{eq:varphiR} and Eq.~\eqref{eq:varphi2level} to be valid within the Beer-Lambert regime.

  

To perform atomic spectroscopy, we release the atomic samples from the optical lattices and immediately apply a probe pulse of duration \(\tau_{\rm p} = 20~\mu\text{s}\). This probe pulse is generated by an external cavity diode laser (Moglabs CEL) with its frequency referred to rubidium vapor saturation spectroscopy~\cite{foot:LFNB,foot:LFN}. The probe's polarization is aligned along \({\bf e}_x\), and the local intensity is set to 3~mW/cm\(^2\), corresponding to a saturation parameter \(s_0 \approx 1\) for steady-state \(\pi\)-excitation~\cite{Steck2003}. These parameters are chosen with the understanding that phase-angle spectroscopy is resilient to power broadening, as discussed in Sec.~\ref{sec:cPhase2level} and to be demonstrated in Fig.~\ref{Fig4}.

In repeated experiments, we scan the probe detuning \(\Delta = \omega - \omega_{eg}\) across the D2 hyperfine \(F=2 \to F'=3\) transition, acquiring up to 9 holograms at each detuning step. A typical reduced hologram, \(\delta I = I' - I\), taken on resonance with \(\Delta \approx 0\), is shown in Fig.~\ref{Fig2}c.


\subsection{Results}

With the $\{I',I\}-$hologram set and the pre-characterized $E_r$, We follow the procedure outlined in Sec.~\ref{sec:principle} to reconstruct $E_s$ using the Gaussian-decomposition method. To initiate the nonlinear optimization (Eq.~(\ref{eq:loss})), we first directly propagate $E_{s}^{(0)}=\delta I/E_r^*$ to an estimated atomic sample plane $\bar z_A$. As evident from Eq.~(\ref{eq:dI}), with this step $E_s(\bar z_A)$ can be focused, albeit being contaminated by a weak, defocused background associated with the ``twin'' term $E_{s,{\rm twin}}$  (Eq.~\eqref{eq:twin}), as well as a typically even weaker ``dc'' $|E_s|^2/E_r^*$ term. Nevertheless, the preliminary $E_{s}^{(0)}(\bar z_A)$ image in Fig.~\ref{Fig2}b helps us to initialize the $\{c_j, {\bf P}_j\}$-parameter and compute $E_G^{(0)}$,  as in Fig.~\ref{Fig2}d, with which we evaluate the expected (reduced) hologram $\delta I_e^{(0)}=E_r^*E_G^{(0)}+E_r{E_G^{(0)}}^*+{|E_G^{(0)}|}^2$ in Fig.~\ref{Fig2}e.

As indicated by Eq.~(\ref{eq:loss}), minimizing $\mathcal{L}_{\rm total}$ requires the simultaneous minimization of both $\mathcal{L}_H = \sum |\delta I_e - \delta I|^2$ and a set of constraints $\mathcal{C}_{\mu}(\varphi(z_A))$ in $\mathcal{L}_A$, which correspond to the expected distribution of $\varphi_A$. Here we impose spatial constraints (Appendix~\ref{sec:Ssupport}) and phase-uniformity constraints (Appendix~\ref{sec:pUniform}). The detuning $\Delta$, subject to frequency-lock uncertainties in saturation spectroscopy~\cite{foot:LFNB}, is not known a priori. 
Instead of fixing $\beta(\Delta)$ (Appendix~\ref{sec:betaFix}), we retrieve $\beta$ from single-shot holograms (Sec.~\ref{sec:comspec}), after refining the atomic sample planes $z_A$ from their single-shot values using multiple holograms at different detunings $\Delta$~\cite{Zhao2022b}. This also helps verify Eq.~\eqref{eq:varphi2level} for free-flying atomic samples.



With $n_A = 3$ Gaussians, the final Gaussian approximation $\langle E_s \rangle_G$ to $E_s$ is presented at $z = \bar{z}_A$ in Fig.~\ref{Fig2}f for each sample, and the optimized $\delta I_e$ is shown in Fig.~\ref{Fig2}g. Subsequently, using Eq.~\eqref{eq:halfstep}, we compute $\tilde{E}_s$, which is depicted at $z = \bar{z}_A$ in Fig.~\ref{Fig2}h. Increasing the number of Gaussians to $n_A = 4$ did not lead to significant improvements in $\delta I_e$ nor $\delta E_G$ (as per Eq.~\eqref{eq:dE}). Therefore, we conclude that $n_A = 3$ is sufficient for decomposing the $E_s$ of the small, spherical samples (Sec.~\ref{subsec:optimize}).


With $\tilde E_s\approx E_s$ and $E_r$ at hand, in Fig.~\ref{Fig2}i we evaluate $\varphi_A$ for all the samples at the fixed, ``common'' $z=\bar z_A$, and plot the OD and $\phi$ images. The array of samples are axially displaced by tens of micrometers. Here, with $\Delta\ll\Gamma$ we expect $\phi\ll {\rm OD}$ according to Eq.~(\ref{eq:varphi2level}), which is satisfied for arrays of samples enclosed by dash-lined boxes only. For these samples, their $z_A$ are close enough to $\bar z_A$ in the display. For other samples, the out-of-focus images mixes ${\rm OD}$ with $\phi$, as expected.



\subsubsection{Sample position localization}\label{sec:zLoc}

\begin{figure}[htbp]
\centering
\includegraphics[width=\linewidth]{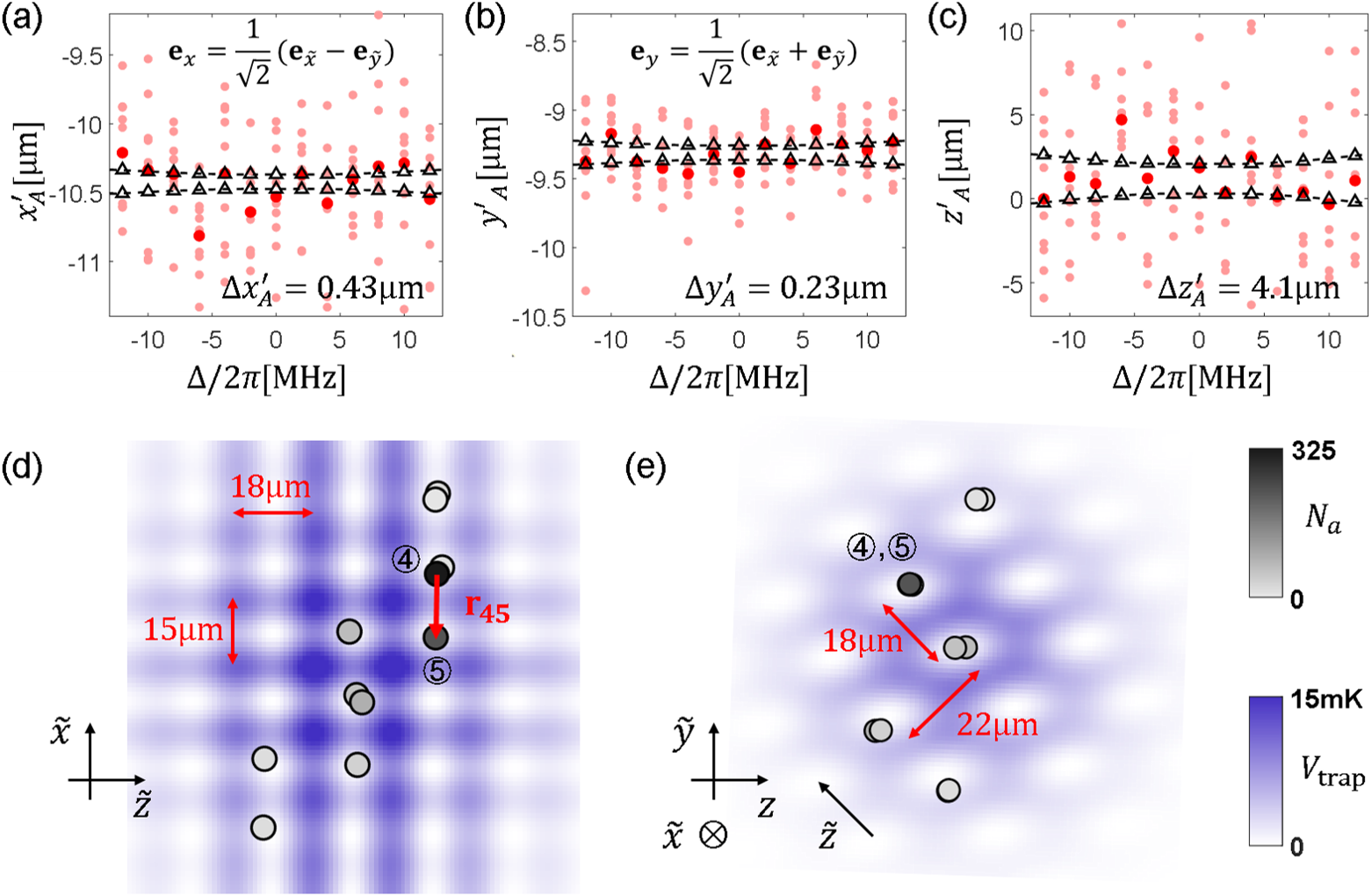}
\caption{(a)(b)(c): Relative displacement ${\bf r}_{45}=(x'_A,y'_A,z'_A)$ between two samples marked in (d)(e) (and Fig.~\ref{Fig2}h) estimated with single-shot $\delta I$ using Eq.~\eqref{eq:paU} criterion. The single-shot estimation at different probe detunings are plotted with red dots. Black triangles indicate the photon-shot-noise limits according to Eq.~\eqref{eq:CRB}. The standard deviations ${\Delta x'_A}$, ${\Delta y'_A}$, ${\Delta z'_A}$ in the figures are averaged over all data at detunings between $-12 {\rm MHz}\leq\Delta/2\pi\leq 12{\rm MHz}$.  (d)(e): Visualization of the sample locations in the 3D lattice. Here the locations are collectively decided by the phase angles~\cite{Zhao2022b} using Eq.~(\ref{eq:varphi2level}) with the $4{\rm MHz}\leq|\Delta|/2\pi\leq 12{\rm MHz}$ data. The atom number $N_a$ is estimated by Eq.~\eqref{eq:varphi2level} with $\xi(s)\propto 1/(1+s)$~\cite{Steck2003}. The circles represent the atomic samples with radius representing the sample sizes. The sizes are larger than $\Delta x'_A, \Delta y'_A$, as well as the collectively-improved $\Delta z'_A \leq 1.5 \rm{\mu m}$. 
}
\label{Fig3}
\end{figure}



With the $E_s$ decomposition based on single-shot $(I',I)$ data, we obtain a list of $z_A$ and the low-resolution $\langle\varphi_A\rangle_G$ for each samples. Using $|\langle\varphi_A\rangle_G|$ in the $x-y$ plane, the center-of-mass locations  ${\bf r}_A=\{x_A,y_A,z_A\}$ are retrieved. As being outlined in Sec.~\ref{sec:principle}, for the $z_{\sigma}\ll z_A,z_H-z_A$ samples here, we  refine the $z_A$ estimation simply by reconstructing $\tilde E_s$ with Eq.~\eqref{eq:halfstep}, and then re-applying the phase-angle uniformity criterion (Eq.~\eqref{eq:paU}). Depending on the $E_s$ complexity and the $n_A$ exploited for the decomposition ($n_A=3$ here), the $\tilde E_s$-refinement generally leads to slightly improved $z_A-$estimation. Here, for samples with substantial ${\rm OD}$ and $\phi$, such as the $\Circled{4},\Circled{5}$ samples for Fig.~\ref{Fig2}(a-c) data, this refinement corrects small $\Delta-$dependent errors in $z_A$ by $2\sim 5~\mu$m.


To estimate the accuracy of the atomic sample localization, ideally we would like to analyze the fluctuation of the individual position values retrieved from repeated measurements. However, our sparse lattice is not stable enough. The lattice centers translate together by $1\sim2~\mu$m from shot-to-shot, likely caused by pointing-instability of the lattice beams (Fig.~\ref{Fig2}a). While the fluctuation hardly impacts our axial, $z_A$-localization analysis, its suppression is required for analyzing the statistics on the more precisely located $x_A,y_A$ values. To suppress the position fluctuation, in Fig.~\ref{Fig3}a-c we instead analyze the relative displacement between two samples. Here, for the single-shot analysis, we choose samples with largest atom numbers, marked as "$\Circled{4}$, $\Circled{5}$" according to Fig.~\ref{Fig2}h, and denote ${\bf r}_{45}=(x'_A,y'_A,z'_A)$. In Appendix~\ref{sec:Fisher} we evaluate the shot-noise-limited precision to the relative displacement as the classical Cramer-Rao bounds with $N_s=\sum_{x,y} |E_{s,A}|^2$ photons (Eq.~\eqref{eq:CRB}). We find the transverse $\Delta x'_A, \Delta y'_A$ are about a factor of 4$\sim$7 larger than the $60~$nm-level photon shot-noise limit (Eq.~\eqref{eq:CRB}), which are likely due to shot-to-shot relative motion between the pair of samples. On the other hand, the 4~$\mu$m axial $\Delta z'_A$ is a factor of 4 larger than the photon shot-noise limit (Eq.~\eqref{eq:CRB}). That the single-shot $\Delta z_A$ is quite far from the shot-noise-limit is likely related to non-optimal data analysis in presence of imaging noises. Nevertheless, both the transverse $\Delta x'_A$, $\Delta y'_A$, and the axial $\Delta z'_A$, are already all well-below the diffraction limits set by $\lambda/{\rm NA}=3~\mu$m and $2\lambda/{\rm NA}^2=23~\mu$m, respectively.



Therefore, as in Fig.~\ref{Fig3}c, we are able to ``super-resolve'' $z_A$ for each sample with $\sim 10^{2}$ atoms even using a single-shot $(I',I)-$data with $\tau_{\rm p}=20~\mu$s exposure. The accuracy can be improved further by exploiting the spectroscopic constraint with the full $\{I',I\}$ data set~\cite{Zhao2022b} collectively. In particular, by enforcing the ${\rm tan}\beta\propto \Delta$ relation required by Eq.~(\ref{eq:varphi2level}) for all the data (except for $|\Delta|/2\pi<4~{\rm MHz}$ data, where the radiation pressure effect is most predominant, as following), our $z_A$ estimation is substantially improved. For the $\Circled{4},\Circled{5}$ samples, the $\Delta z_A$ are reduced to below one micrometer. Comparing with single-shot $z_A$ localization as in Fig.~\ref{Fig3}c, the refined $z_A$ typically involves 2-3~$\mu$m additional corrections. These corrections helps to ensure that the samples are more consistently located within dark centers of the 3D lattice, as illustrated in Figs.~\ref{Fig3}(d)(e).





\subsubsection{Complex-valued spectroscopic imaging}\label{sec:comspec}
\begin{figure}[htbp]
\centering
\includegraphics[width=\linewidth]{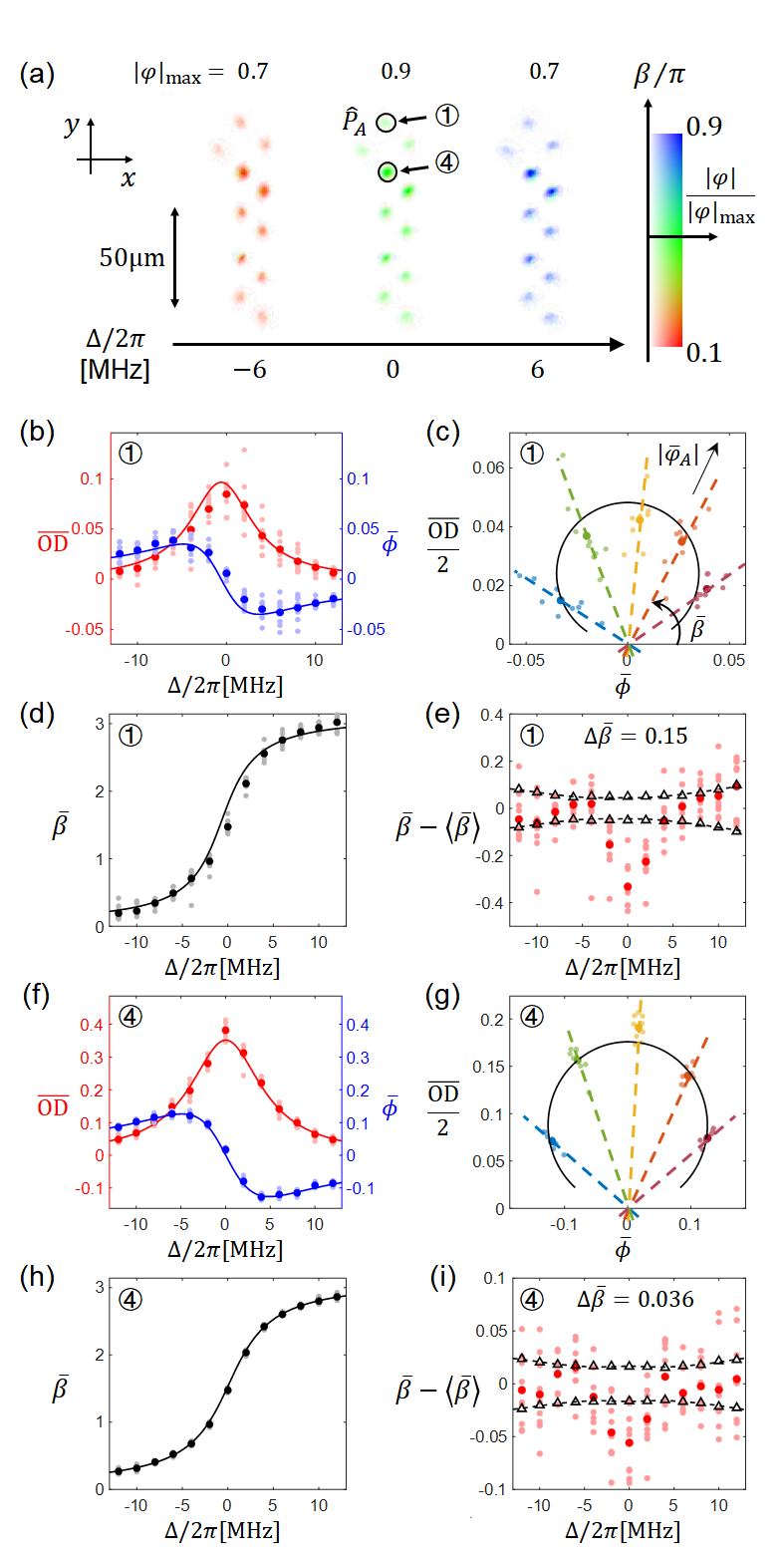}
\caption{ Complex-valued spectroscopic imaging. (a): Color domain plot of $\varphi_A$ with all the samples projected to the $z=\bar z_A$ plane. (b)(f): Average $\overline{\rm OD}$, $\overline{\phi}$, within $\hat{P}_A$ marked in (a), for "1,4" samples respectively. Solid (transparent) dots are average (individual) values of nine measurements. Solid lines are fitted according to Eq.~\eqref{eq:varphi2level}. (d)(h): The phase-angle spectroscopy. Notice in (h) the 9 sets of data overlap closely. (c)(g): $\overline{{\rm OD}}/2$-$\overline{\phi}$ phasor plot at $\Delta/2\pi=-6~{\rm MHz},-2~{\rm MHz},0~{\rm MHz},2~{\rm MHz}$ and $6~{\rm MHz}$, in red, orange, yellow, green, and blue respectively. The black lines are according to the (b)(f) fits. (e)(i): Residuals of $\bar{\beta}$ from the (d,h) fits. Black triangles and dashed lines give photon-shot-noise limit~\cite{Wang2022b} (Eq.~\eqref{eq:db}). The phase-angle uncertainty $\Delta \bar{\beta}$ of $0.15$ and $0.036$ are obtained by averaging the shot-to-shot variations at each $\Delta$~\cite{foot:exc}. The values are $50\%\sim100\%$ larger than the corresponding photon-shot-noise limit (Eq.~\eqref{eq:db})~\cite{foot:LFN2}.
}
\label{Fig4}
\end{figure}


With the list of $z_A$ and $\langle \varphi_A\rangle_G$ for each sample, we can proceed to evaluate $\tilde E_{s,A}$ with Eqs.~(\ref{eq:halfstep})(\ref{eq:EsApp}) and to approximately obtain $\varphi_A$ with diffraction-limited resolution, according to Eq.~(\ref{eq:varphiR}).
However, as discussed in Appendix~\ref{sec:Fisher}, the Eq.~(\ref{eq:db}) shot-noise-limit to the phase-angle $\beta$ is reached with fixed $z_A$. Without the knowledge, the phase-angle uncertainty $\Delta \beta$ is compromised by a factor of $1/\sqrt{2}$, at least. Therefore, we re-optimize the Gaussian decomposition, using the $\{I,I'\}$-set, collectively improved $z_A$ as prior knowledge. The phase-angle spectroscopy to be discussed in the following is based on such re-optimized single-shot Gaussian decomposition (Fig.~\ref{Fig3}(d,e)).

We now evaluate $\varphi_A=\varphi(z_A)$ using Eqs.~\eqref{eq:varphiR}\eqref{eq:EsApp}. The color-domain plots in Fig.~\ref{Fig4}a show $\varphi_A$, with phase angle $\beta$ indicated by color and modulus $|\varphi_A|$ by brightness. In contrast to Fig.~\ref{Fig2}(i), $\varphi(z_A)$s from various samples are “projected" to $z=\bar z_A$ from different $z_A$ planes. In the plots we observe consistent phase-angle $\beta$ at the same detunings $\Delta$, as anticipated. It is also observed that the same sample may appear slightly different at various detunings. This slight distortion in shape is primarily attributed to the lensing effect as the probe passes through the microscopic samples.


To further demonstrate aspects of the complex-valued spectroscopy, in Figs.~\ref{Fig4}(b)(f) we choose $\Circled{1},\Circled{4}$ samples and plot $\overline{\rm OD}=2{\rm Im}[\bar\varphi_A]$, $\overline{\phi}={\rm Re}[\bar\varphi_A]$ as a function of the probe detuning $\Delta$. Here $\bar\varphi_A$ is averaged over the sample area $\hat P_A$ as suggested in Fig.~\ref{Fig4}a. The same data is further presented in Figs.~\ref{Fig4}(c)(g) as 2D phasor plots, and in Figs.~\ref{Fig4}(d)(h) in terms of phase-angle spectroscopy. 

In Figs.~\ref{Fig4}(b)(f) we see data from the repeated single-shot measurements fluctuate substantially. A closer look suggests there are strong correlations between the $\overline{\rm OD}$ and $\bar\phi$ fluctuations, which are more clearly seen in Figs.~\ref{Fig4}(c,g): The repeated measurements mostly spread the data points along the radial $|\varphi_A|$ direction instead of along $\beta$. These fluctuations are associated with the quite-expected atom number fluctuations in repeated measurements, which strongly affect the modulus $|\varphi_A|$ but hardly affect $\beta$, according to Eq.~(\ref{eq:varphi2level}).


The stability of $\beta$ is clearly evident in the phase-angle plot shown in Fig.~\ref{Fig4}(d)(h), where fluctuations are significantly suppressed. Using ${\rm tan}\bar \beta = -2\Delta/\tilde{\Gamma}$~\cite{foot:LFNB}, based on Eq.~(\ref{eq:varphi2level}), we fit the data while excluding the points in the range $-4~{\rm MHz} < \Delta/2\pi < 4~{\rm MHz}$. The fit is weighted by the amplitude $|\bar \varphi_A|$. Consistent extraction of $\tilde{\Gamma}/2\pi \approx 6~{\rm MHz}$ is achieved across all $\Circled{1}$ to $\Circled{11}$ samples. Moderate deviations of up to 0.7~MHz are observed in the fitted values of $\tilde{\Gamma}$, compared to $\Gamma = 2\pi \times 6.1$~MHz. These deviations, which vary from sample to sample, are not fully understood but may stem from inaccuracies in data processing, such as those caused by lensing effects (Fig.~\ref{Fig4}(a)). In contrast, the $\overline{\rm OD}$ and $\bar\phi$ curves in Fig.~\ref{Fig4}(b)(f) exhibit significant power broadening, with Lorentzian fits indicating apparent widths of $8$--$9$~MHz, consistent with the expected broadening at $s_0 \approx 1$.

The phase-angle residuals are shown in Figs.~\ref{Fig4}(e)(i). For all the samples, we consistently see deviation of the $\bar \beta$-data from the 2-level model within $\Delta<\Gamma/2$. As in Appendix~\ref{sec:Doppler}, this deviation is caused by Doppler shifts associated with radiation pressure. In particular, for the $\tau_{\rm p}=20~\mu$s probe with $s_0=1$ on resonance, the atomic sample is expected to be accelerated to a final $\delta v\approx 1$~m/s along ${\bf e}_z$. The average $\delta \omega_{eg}\approx k_0\delta v /2 \sim 2\pi\times 640~$kHz Doppler shift is sensed by our complex spectroscopy (Eq.~\eqref{eq:Doppler}), leading to $\delta \bar \beta=k_0\delta v/\Gamma\approx 0.2$ phase-angle shift. Remarkably, the shift is almost halved for the $\Circled{4}$ sample, as in Fig.~\ref{Fig4}i, which is likely related to reduced radiation pressure~\cite{Bachelard2016} due to attenuation of the probe by the larger sample.  

 
We note that on resonance the radiation pressure is also expected to displace the samples by up to $10~\mu$m along $z$ while transversely induce a $\sim 1~\mu$m spreading by heating. The average $z_A$ displacement during $\tau_{\rm p}$ is $2\sim 3~\mu$m-level at most. The differential displacement between $\Circled{4},\Circled{5}$ samples is below the axial resolution in Fig.~\ref{Fig3}(c). The sub-micron average spreading is smaller than both the sample size $\sigma$ and the $x_{\rm res}=3~\mu$m  transverse resolution, hardly detectable either.



We estimate the phase-angle resolution $\Delta \bar \beta$ using the data statistics from Fig.~\ref{Fig4}(e)(i). Theoretically (Appendix~\ref{sec:Fisher}), for optically thin samples, the resolution $\Delta \bar \beta(\Delta)$ is constrained by $N_s(\Delta)$, which represents the number of coherently scattered photons $E_s$ received by the camera, through the Eq.~\eqref{eq:db} relation. After applying a moderate correction for finite $|\varphi_A|$ as discussed in ref.~\cite{Wang2022b}, the shot-noise-limited $\Delta \bar\beta$ values are evaluated according to Eq.~\eqref{eq:db} and marked with triangle symbols in Fig.~\ref{Fig4}(e)(i). We observe that the values of $\Delta\bar \beta$ closely follow the photon-shot-noise limit for both the $\Circled{1}$ and $\Circled{4}$ samples. By averaging the standard deviations of $\Delta\bar\beta(\Delta)$ at each $\Delta$ in Fig.~\ref{Fig4}, we estimate statistically the single-shot resolution to be $\Delta\bar\beta \approx 0.15$ and $0.036$ for the $\Circled{1}$ and $\Circled{4}$ samples, respectively~\cite{foot:exc, foot:LFN2}. For the $\Circled{4}$ sample, the 36~mrad resolution suggests a frequency resolution of $\Delta\omega=2\pi\times 100~{\rm kHz}$, according to Fig.~\ref{Fig4}h and Eq.~(\ref{eq:varphi2level}) near $\Delta=0$ and $\Delta\omega=\Delta \beta \times \Gamma/2$, within single shot. As demonstrated in Fig.~\ref{Fig4}(e)(i), the frequency resolution is fine enough to resolve the Doppler shift caused by resonant radiation pressure, a notable effect when considering interaction dynamics~\cite{Bachelard2016}. However, to accurately determine the absolute atomic line center in future work, especially with single-shot measurements, then, a careful choice of probe detuning $\Delta$ is required to balance $N_s(\Delta)$, the $\partial_{\Delta}\beta$ sensitivity, and the Doppler shift systematic during the prolonged probe. It is also possible to mitigate the Doppler shift by interleaving shorter probe pulses with cooling and trapping~\cite{Zhao2022b}. Further discussion on resolving Doppler-shifted lines and correcting systematic errors is provided in Appendix~\ref{sec:Doppler}.


\subsubsection{Field sensing}

\begin{figure}[htbp]
\centering
\includegraphics[width=\linewidth]{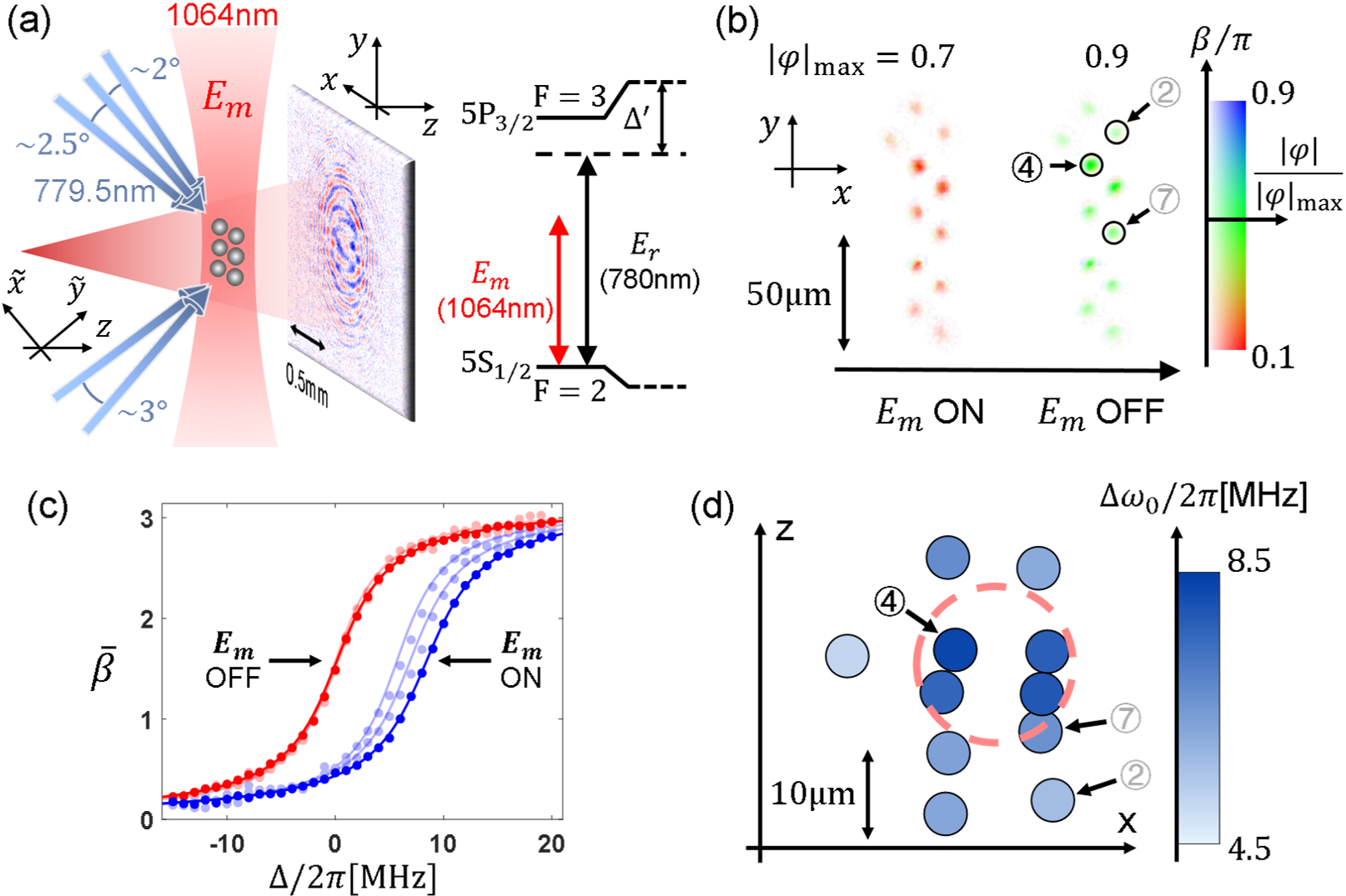}
\caption{Sensing light shifts with a sparse array of laser-cooled atoms. (a): Experimental setup and the level diagram. (b): $\bar z_A$-projected $\varphi_A$-images of the axially displaced samples, with and without $E_m$. The data are processed from $\delta I$ averaged over 9 exposures to enhance the display. The probe detuning is $\Delta/2\pi = 0~{\rm MHz}$~\cite{foot:LFNB} (c): $\hat P_A$-averaged phase angle spectra of "2,4,7" samples marked in (b). The red (blue) curves are for $E_m$ off (on). The transparency levels are according to the markers in (b). (d): $x-z$ ``top'' view of the light shift distribution. The 11 largest-$|\varphi|$ samples are represented by blue disks. The dash-lined circle suggests the estimated $E_m$ full-width-half-maximum focal spot in the $x-z$ plane.
}
\label{Fig5}
\end{figure}

With the spectroscopic images of atomic arrays, it is possible to detect local perturbation to the $|g\rangle-|e\rangle$ transition frequency $\omega_{eg}$ in 3D with high spatial resolution.  In this section, we demonstrate this field-sensing capacity by measuring the transition light shift induced by a  ``dressing'' laser. The schematic setup is displayed in Fig.~\ref{Fig5}a. Right before the imaging exposure, a dressing laser $E_m$ with a Gaussian radius of $w_m\approx 20~\mu$m, at $\lambda_m=1064~$nm and with power $P_m\approx 1~$W, is turned on. This leads to Stark shift $\delta \omega_{eg}\propto {|E_m|}^2$ of about 6~MHz, as been observed in Fig.~\ref{Fig5}b, with the $\varphi_A$ images with quantitative agreements. 

Taking a closer look at Fig.~\ref{Fig5}b, since the dressing beam is not uniform, the 3D-distributed samples experience slightly different $\delta\omega_{eg}$ shifts.  In Fig.~\ref{Fig5}c we plot $\bar \beta={\rm arg}(\bar \varphi_A)$ for three samples, labeled as $\Circled{2}$, $\Circled{4}$, $\Circled{7}$ in Fig.~\ref{Fig5}b. For the convenience of display, we encode the sample index with the degree of transparency for the $\bar \beta$ curves. We clearly see that while in absence of dressing the $\bar \beta$ curves closely overlap (the red curves), the three $\bar \beta$ curves disperse apart when $E_m$ is on (blue curves). The difference is most pronounced near the shifted $\omega_{eg}$ center, with $\Delta/2\pi \approx 6~$MHz, where $\bar \beta$ regains the highest $\Delta$-sensitivity of $\partial_{\Delta}\beta\sim 2/\Gamma$ according to Eq.~(\ref{eq:varphi2level}). By analyzing the spectra for all samples, we are able to resolve $\delta \omega_{eg}$ in 3D with micro-meter resolution. The spatial-dependent light shift is most conveniently displayed in the $x-z$ plane where the dressing beam is in focus. For the purpose, the vertical view of $\Delta\omega_{eg}$ is plotted in Fig.~\ref{Fig5}d for our eleven samples with significant $|\varphi_A|$. 

The field-sensing capacity demonstrated in this section shares the same frequency resolution with that is discussed in the last section. In particular, according to Eq.~(\ref{eq:varphi2level}) near the shifted $\omega_{eg}$ resonance, we again have $\Delta (\delta \omega_{eg})/2\pi \approx 100~{\rm kHz}$ to sense the shifted line. We emphasize this resolution is achieved by $\sim 10^2$ atoms within $\tau_{\rm p}=20~\mu$s, with micrometer 3D resolution.



\section{Discussions}\label{sec:conclusion}

\subsection{Aberrations}\label{sec:aberr}
So far, our holographic imaging scheme assumes aberration-free propagation of $E_{r}$, $E_s$ between the sample plane $z_A$ and hologram plane $z_H$. Practically, the $E_{r,s}$ wavefronts need to be optically relayed from the ultrahigh vacuum to the place where the digital camera sensor conveniently locates. For high-NA imaging, aberration becomes unavoidable when the required volume-of-view is large.

To describe the imperfect imaging, we consider both  $E_r$ and $E_s$ as in Fig.~\ref{Fig1}a to be relayed from inside the ultrahigh vacuum where the actual light-atom interaction takes place. We use a wavefront propagator $U_{\rm img}$ to generally represent the linear transformation of wavefronts from the in-situ $z=z_{V,A}$ to $z=z_A$,
\begin{equation}
    \begin{aligned} 
        E_r(x,y,z_A)&={U}_{\rm img} E_{r}(x_V,y_V,z_{V,A}),\\
        E_s(x,y,z_A)&={U}_{\rm img} E_{s}(x_V,y_V,z_{V,A}).
    \end{aligned}\label{eq:OTF1}
\end{equation}
For imperfect imaging, the fundamental way to eliminate the associated imaging errors is to fully characterize the propagator $U_{\rm img}$. Then, by resolving $E_{r,s}(z_{V,A})$ through Eq.~\eqref{eq:OTF1} to apply our prior knowledge about the samples in situ. Practically, even with advanced computing power, the modeling of diffraction-limited $E_r,E_s$ propagation across the large-diameter lens-array can still be prohibitively resource-demanding. On the other hand, the wavefront propagation is greatly simplified by Gaussian decomposition~\cite{Heller1975, Harvey2015,Ashcraft2020}. The well-developed technique is naturally adapted to the Gaussian-assisted holography scheme in this work to potentially enable aberration-free imaging of sparse samples over unprecedentedly large imaging volumes, only limited by the finest hologram fringes that can be captured by the camera~\cite{Sobol2014a}.

To fully characterize $U_{\rm img}$ for the large-diameter lens-arrays~\cite{Alt2002,LI2018}, and to invert Eq.~\eqref{eq:OTF1} with Gaussian-decomposition, are all beyond the scope of this work. Instead, we take advantage of the small volume-of-view  (Fig.~\ref{Fig2}) to address minor, aplanatic aberrations. This is achieved by employing a point-spread function~\cite{Altuntas2021} $O(x,y)$ through in situ optical measurements. The procedure is detailed in Appendix~\ref{sec:aberrApp}.

\subsection{Speckles}
In addition to aberrations, the imaging optics may introduce speckle noise into the wavefronts through multiple reflections and even by small dusk particles on the optical surfaces. Since the probe beam for atomic spectroscopy is highly monochromatic, the speckle noises are difficult to avoid.  

We denote $\delta E_r$ and $\delta E_s$ to be speckle fields generated by unwanted reflections and scatterings across the imaging optical path. We assume that the perturbations are static enough for their precise characterization through multiple measurements. With Gerchberg-Saxton iterations~\cite{Wang2022b}, $\delta E_r$ is fully characterized in this work (Fig.~\ref{Fig1}b). When speckle noises in $E_{r,s}$ are moderate, as in this work, their presence hardly affect our prior knowledge about $\varphi_A$ for the $E_s$ retrieval (Sec.~\ref{sec:principle}). But the presence of $\delta E_r$ does affect the $\varphi_A$ retrieval with Eqs.~\eqref{eq:varphiR}, particularly since substantial amount of the speckle noises $\delta E_r$ may be generated by imaging optics {\it after} the light-atom interaction, not ``seen'' by atoms at $z=z_A$. To minimize the $\delta E_r$ impact, one should locate the sources of $\delta E_r$ with digital propagation of $E_r$, and by removing the speckles to recover the actual $E_r$ attending the light-atom interaction.  The process of characterizing the speckle sources across the imaging path is therefore similar to characterizing $U_{\rm img}$ discussed above. If all the speckle scatterings in $\delta E_{r,s}$ are static enough, the procedure would help to fully suppress the associated noises in the $\varphi_A$ retrieval.

In this work, we take a simpler approach to approximately suppress the speckle noise to the complex-valued spectroscopic imaging. In particular, during the Eq.~\eqref{eq:varphiR} evaluation of $\varphi_A$, we simply filter away the apparent speckles  from $E_r$. This is achieved by digitally focusing $E_r$ back to $z=0$ and only choose a small region around the diffraction-limited component, before propagating forward to $z=z_A$ for the $\varphi_A$ evaluation. In other words, we assumes $E_r$ ``seen'' by atoms is speckle-less, which is a good approximation in our setup, considering that comparing to the multiple-stage-relayed large-diameter imaging optics, the $E_r$ is focused to atomic samples with simpler, smaller-diameter optics~\cite{Alt2002}.

\subsection{Summary}
From precision spectroscopy~\cite{Bons2016,Marti2018,Li2020} to quantum simulation~\cite{Gross2017}, precise optical imaging~\cite{Bakr2009,Sherson2010a,Okuno2020} is crucially important for advancing ultracold atomic physics research at frontiers. The prevailing imaging method in these researches is to directly record fluorescence~\cite{Nelson2007, Bakr2009, Sherson2010a,
Barredo2018,Eliasson2020, Legrand2024}, absorption~\cite{Spielman2007, Hung2011}, and phase shift~\cite{Higbie2005}.  While holographic imaging techniques are rapidly developing for applications in other fields~\cite{Isikman2011, Greenbaum2012,DeHaan2020,Park2021}, their utilities in atomic physics are rare~\cite{Kadlecek2001,Turner2005,Sobol2014,Smits2020,Altuntas2021,Wang2022b}. In Sec.~\ref{sec:challenge} we highlighted major challenges of quantitative holographic imaging for precision spectroscopic measurements of cold atoms. We emphasized that the holographic measurement is highly useful for spectroscopy with cold atomic ensembles to overcome intrinsic atom-number and interaction-strength fluctuations (Sec.~\ref{sec:normf}), fundamentally efficient (Sec.~\ref{sec:spec1}), and offers the unique opportunity of single-shot phase-angle readouts with power-broadening resilience (Sec.~\ref{sec:cPhase2level}). If the challenges can be addressed, then, by extending the spectroscopy data to complex numbers and by enabling diffraction-limited 3D resolution (Sec.~\ref{sec:multiI}), the holographic method would help to unlock highly exciting opportunities associated with precise spectroscopic imaging and sensing with cold atomic ensembles on strong optical transitions.


In this work, motivated by the exciting prospect of complex-valued 3D spectroscopic imaging, we develop a systematic approach to address key challenges in holographic imaging of cold atoms (Sec.~\ref{sec:challenge}). Comparing with {\it e.g.} biological applications~\cite{Ou2013,Park2021a}, our choices of schemes and diversities in holographic measurements are severely constrained by aspects of cold-atom setups. Our philosophy is to fully characterize the imaging setup and to efficiently utilize all the available knowledge from single-shot inline holography. The efficient application of prior knowledge is numerically assisted by Gaussian beam propagation, a simple application of the well-established Gaussian-decomposition technique~\cite{Heller1975,Kong2016ThawGauss,Harvey2015, Ashcraft2020,Worku2020}. With the method,  we successfully demonstrate complex-valued spectroscopic imaging of axially displaced microscopic $^{87}$Rb samples in a sparse lattice with micrometer-level 3D resolution. Atom numbers and transition frequencies are simultaneously inferred on each lattice site, within single shot. We achieve hundred-kHz-level single-shot frequency resolution out of the $\Gamma/2\pi=6~$MHz natural linewidth, with merely hundreds of atoms, a result that appears extremely difficult to realize with traditional spectroscopy~\cite{Sansonetti2011,Li2020} in presence of the atom number and interaction strength uncertainties.

Not discussed in this paper is the observation of atomic shot noise~\cite{Hung2011,Altuntas2021}, which is barely supported by but nevertheless observed with our ${\rm NA}\approx 0.26$ setup. In addition, the imaging volume-of-view in this work is limited by the sample distribution itself to be within a $\sim$100$~\mu$m distance (Also see Appendix~\ref{sec:large}). 
We clarified in Sec.~\ref{sec:aberr} that the technique can be extended to fully suppress wavefront aberrations for wide-field 3D imaging of sparse atomic samples. The development may enable wideband, spectroscopic sensing of 3D potentials~\cite{stamperkurn,Diego2023} with unprecedented volume-of-view. Other than probing single-atom spectroscopic responses, our method also supports the study of resonant dipole interactions by simultaneously measuring the strength, direction and phase angle of the collective emission~\cite{Zhu2016, Ji2023}. As such, we also expect the complex-valued spectroscopic imaging method to facilitate efficient characterization and quantum engineering of advanced light-atom interfaces~\cite{Bons2016,Rui2020,Deb2021, Lu2022,Srakaew2023}.

\section*{Acknowledgements}
We are grateful to Professor Yuan, Haidong for valuable discussions and insightful suggestions, and to Mr.~Wang, Dong for numerical simulations and for preparing Figure~\ref{FigSat}. This work is supported by the National Key Research Program of China (Grant No.~2022YFA1404204), the National Natural Science Foundation of China (NSFC, Grant No.~12074083), and the Original Research Initiative at Fudan University.

\appendix




\section{Cramer-Rao bounds to position and phase localization by inline holography}\label{sec:Fisher}
In this section, we consider the estimation of bulk properties of the atomic sample from the hologram $I,I'$ specified in Eq.~(\ref{eq:dI}). The basic setup is illustrated in Fig.~\ref{Fig1}a. The bulk properties of interest include the 3D atomic central locations ${\bf r}_A=(x_A,y_A,z_A)$ and the total amplitude of the forward scattering $E_s$. To facilitate the derivations, we introduce five $\{\theta_j\}=\{ x_{A}, y_{A}, z_{A}, \beta_{r,{\rm off}}, \beta_{i,{\rm off}}\}$ parameters. The last two parameters are the real and imaginary parts of an auxiliary complex phase offset in
\begin{equation}
\tilde E_s=e^{i\beta_{\rm off}} \hat U(z_H- z_A)\left(E_r(z_A) (e^{i \varphi_A}-1)\right).\label{eq:Estilde}
\end{equation}
For example, with $|\varphi_A|\ll 1$, then $\beta_{r,{\rm off}}$ offsets our estimation to phase-angle $\beta={\rm arg}[\varphi_A]$, and $\beta_{i,{\rm off}}$ offsets $|\varphi_A|$.

The hologram $I$ is accordingly re-modeled as
\begin{equation}
\tilde I =\left|E_r(z_H)+ \tilde E_s\right|^2
\end{equation}
so that $I'=\tilde I|_{\beta_{\rm off}\rightarrow 0}$. We regard $p(x,y)=\tilde I(x,y)/\sum_{x,y}\tilde I(x,y)$ as a 2D probability distribution of a single photon received by the camera.  The Fisher information matrix is accordingly expressed as
\begin{equation}
  \begin{aligned}
F_{j k}^{(0)}&\equiv \left(\sum_{x,y}\frac{1}{ p(x,y)}\partial_{\theta_j} p(x,y) \partial_{\theta_k} p(x,y)\right)_{\beta_{\rm off}\rightarrow 0},
\\
&\approx \left(\frac{\sum_{x,y}\frac{1}{ \tilde I(x,y)}\partial_{\theta_j} \tilde I(x,y) \partial_{\theta_k} \tilde I(x,y)}{\sum_{x,y}  \tilde I(x,y)}\right)_{\beta_{\rm off}\rightarrow 0}.\label{eq:fisher}
  \end{aligned}
\end{equation}
In the second line of Eq.~(\ref{eq:fisher}) we assumed the total flux $\sum_{x,y}\tilde I(x,y)$ received by the camera to be $\theta_j$-independent, which is sufficiently satisfied for moderately-dense atomic samples in the defocused-twin regime (Appendix~\ref{sec:dftwin}) with $z_{\sigma}\ll z_A, z_H-z_A$ (Fig.~\ref{Fig1}a), so that $|E_s|\ll |E_r|$ on the camera plane as to be assumed in the following. With $\partial_{\theta_j}\tilde I \partial_{\theta_k}\tilde I\approx |E_r|^2(\partial_{\theta_j} \tilde E_s^*\partial_{\theta_k}\tilde E_s+c.c)+ (E_r^{*2}\partial_{\theta_j} \tilde E_s\partial_{\theta_k}\tilde E_s+c.c.)$, the second term also vanishes upon the 2D integration, due to the substantial wavefront mismatch between $E_r$ and $E_s$. Equation~(\ref{eq:fisher}) is accordingly simplified into
\begin{equation}
F_{j k}^{(0)}\approx   \frac{\sum_{x,y}\left(\partial_{\theta_j} \tilde E_s^* \partial_{\theta_k} \tilde E_s +c.c. \right)_{\beta_{\rm off}\rightarrow 0}}{\sum_{x,y}  |E_r|^2}.\label{eq:fisher2}
\end{equation}

The Fisher information matrix by Eq.~(\ref{eq:fisher2}) is associated with single-photon detection on the camera. For imaging cold atoms, we are more interested in Fisher information associated with photons scattered by the samples~\cite{Lye2003}. A change of normalization leads to
\begin{equation}
F_{j k}=\left(\frac{\sum_{x,y}\partial_{\theta_j} \tilde E_s^* \partial_{\theta_k} \tilde E_s +c.c.}{\sum_{x,y}  |\tilde E_s|^2}\right)_{\beta_{\rm off}\rightarrow 0}.\label{eq:fisher3}
\end{equation}
Interestingly, Eq.~(\ref{eq:fisher3}) suggests Fisher information for holographic measurements upon a single $E_s$-photon detection is equal to half of the quantum Fisher information~\cite{Braunstein1996}  for the $E_s$-photon wavepacket. This 50\% loss of information is generally associated with heterodying measurements~\cite{Sobol2014}. 

From a single photon wavepacket to $N_s=\sum_{x,y}|E_s|^2$ photons detected by the camera in an uncorrelated manner, we have $F_{j k}(N_s) = N_s F_{j k}$ according to Eq.~(\ref{eq:fisher3}) to limit the resolutions of $\{\theta_j\}$ measurements with the associated Cramer-Rao bounds. 
With Eq.~(\ref{eq:Estilde}), evaluation of Eq.~(\ref{eq:fisher3}) for the $\beta_{r,{\rm off}}$, $\beta_{i,{\rm off}}$ differentiation is straightforward. With the free-space propagator $\hat U(z_H-z_A)$ by Eq.~(\ref{eq:asm}) to be $x_A,y_A$ independent, the translational invariance suggests
\begin{equation}
\begin{aligned}
\partial_{x_A}\tilde E_s=-\partial_x \tilde E_s,\\
\partial_{y_A}\tilde E_s=-\partial_y \tilde E_s.\label{eq:pxy}
\end{aligned}
\end{equation}

For the differentiation of the $z_A-$parameter in Eq.~(\ref{eq:fisher3}), we rewrite
\begin{equation}
\tilde E_s=e^{i\beta_{\rm off}} \hat U(z_H- z_A)\left(\left(\hat U(z_A) E_r(0)\right) (e^{i \varphi_A}-1)\right).\label{eq:Estilde2}
\end{equation}
Here the first propagator $\hat U(z_H-z_A)$ acts on $E_s(z_A)=E_r(z_A)(e^{i\varphi_A}-1)$. The second $\hat U(z_A)$ acts on $E_r(0)$ only. In the $z_{\sigma}\ll z_A$ (Fig.~\ref{Fig1}a) limit, the $E_r(z_A)$ ``seen'' by the $\sigma-$sized atomic sample can be approximated by plane wave. For on-axis samples we therefore find 
\begin{equation}
\partial_{z_A}\tilde E_s\approx i(-\sqrt{k_0^2+\partial_x^2+\partial_y^2}+k_0)\tilde E_s.\label{eq:pz}
\end{equation}

With Eqs.~(\ref{eq:Estilde})(\ref{eq:pxy})(\ref{eq:pz}), evaluation of Eq.~(\ref{eq:fisher3}) becomes straightforward for specific imaging setup, $E_r-$illumination, and atomic samples. To provide the Cramer-Rao bounds with a simple example, in the following we consider $E_r$ to propagate along the $z-$axis, $E_s(z_A)$ to be the forward scattering by a Gaussian shaped sample located on the $z-$axis with a density distribution of $\rho_c=\rho_0 e^{-(x^2+y^2)/\sigma^2}$. We assume $\varphi_A=k_0\rho_c\alpha(\omega)\ll1$ so that the $E_s$ amplitude simply follows the $\rho_c$ profile, while $\sigma> \lambda/{\rm NA}$ so that $E_s$ is almost fully captured by the camera. By normalizing $x_A,y_A$ with $\sigma$, and $z_A$ with $z_{\rm \sigma}$, the rescaled Fisher information matrix with $N_s$ scattered-photon detection is
\begin{equation}
  F(N_s)=N_s
    \left[ {\begin{array}{ccccc}
      2 & 0 & 0 & 0 &0 \\
      0 & 2 & 0 & 0 &0 \\
      0 & 0 & 1 & 1 & 0 \\
      0 & 0 & 1 & 2 & 0 \\
      0 & 0 & 0 & 0 & 2 \\
    \end{array} } \right]. \label{eq:S3}
\end{equation}
Clearly, while $x_A,y_A$ and $\beta_{i,{\rm off}}$ measurements are independent, the axial position $z_A$ is correlated with the phase angle $\beta_{r,{\rm off}}$. By diagonalizing Eq.~(\ref{eq:S3}) and evaluate the Cramer-Rao bounds indirectly using the eigenvector observables, one finds that in the $\varphi_A\ll1$ optically-thin-sample limit,
\begin{equation}
\begin{aligned}
\Delta x_A&=\sigma/\sqrt{2N_s},\\
\Delta y_A&=\sigma/\sqrt{2N_s},\\
\Delta z_A&= z_{\sigma}/\sqrt{N_s/2},\\
\Delta \beta&=1/\sqrt{N_s},\\
\Delta |\varphi_A|&=1/\sqrt{2 N_s}.\label{eq:CRB}
\end{aligned}
\end{equation}
The axial resolution $\Delta z_A$ can be enhanced to $\Delta z_A=z_{\sigma}/\sqrt{N_s}$ limit (Eq.~(\ref{eq:dz}))~\cite{Zhao2022b} by taking multiple spectroscopic measurements to fix $\beta$ first. On the other hand, with the axial position $z_A$ fixed, the phase angle resolution is enhanced to $\Delta \beta=1/\sqrt{2 N_s}$ (Eq.~(\ref{eq:db}))~\cite{Wang2022b}.

Finally, to apply Eq.~\eqref{eq:CRB} for the relative displacement between two samples, such as ${\bf r}_{45}$ in Fig.~\ref{Fig3}, the position variations should be summed. This results in an effective photon number of $N_s = \left( N_{s,4}^{-1} + N_{s,5}^{-1} \right)^{-1}$.


\section{Details of Gaussian decomposition}\label{sec:scheme}
\subsection{Gaussian beam}

Gaussian decomposition is associated with a class of powerful semi-classical wave-propagation techniques~\cite{Heller1975,Kong2016ThawGauss,Harvey2015, Ashcraft2020,Worku2020}. We focus on the free Gaussian-propagation to support evaluation of $E_r, E_s$ across  $z-$planes during the minimization of $\mathcal{L}_{\rm tot}$ in Eq.~(\ref{eq:loss}).

The standard form of a Gaussian beam is expressed as 
\begin{equation}
        G_z({\bf r};{\bf P}')=e^{i({\bf r}_{\perp}-{\bf r}_{\perp}^{0})^T\frac{k_0}{{2\bf q}(z)}({\bf r}_{\perp}-{\bf r}_{\perp}^{0})+i k_0 z+i\varphi_G(z)}. \label{eq:Gz}
\end{equation}
Here ${\bf r}_{\perp}=(x,y)^T$. The complex ${\bf q}(z)$-parameter is defined by 
\begin{equation}
\frac{1}{{\bf q}(z)}=\left[ {\begin{array}{cc}
  \frac{1}{R_x} + i\frac{2}{k_0 w_x^2}  & 0 \\
  0  & \frac{1}{R_y} + i\frac{2}{k_0 w_y^2}, \\
\end{array}}  \right].\label{eq:Gw}
\end{equation}
with
\begin{equation}
  \begin{aligned}
  w_{c}(z)&=w_{c}^0\sqrt{1+\frac{(z-z_c^0)^2}{(z_{c}^R)^2}},\\
  R_{c}(z)&=z-z_c^0+\frac{(z_{c}^{\rm R})^2}{z-z_c^0},
  \end{aligned}
\end{equation}
all decided by the Gaussian waists $w_c^0$ for $c=x,y$. 
The Gouy phase $\varphi_G(z)$ is given by
\begin{equation}    
  \varphi_G(z) =\frac{1}{2}\sum_{c=x,y}{\rm arctan}((z-z_c^0)/z_{c}^R).
\end{equation}
The Rayleigh lengths are defined as
\begin{equation}
  z_{c}^{\rm R}=\pi (w_{c}^0)^2/\lambda \label{eq:Rayleigh}
\end{equation}
for $c=x,y$.

Next, we introduce a rotation matrix $\mathcal{R}({\gamma})$ 
with Euler angles $\gamma=(\theta_1,\theta_2,\theta_3)$, so that  ${\bf k}=(0,0,k_0)\mathcal{R}(\gamma)$ is expressed as
\begin{equation}
  {\bf k}=(-{\rm sin}\theta_2, {\rm sin}\theta_1 {\rm cos}\theta_2,{\rm cos}\theta_1{\rm cos}\theta_2)k_0.
\end{equation}
The rotation angle $\theta_3$ orients the principal axis of the 2D Gaussian profile.  With ${\bf r}=(x,y,z)^T$, we have
\begin{equation}
  G({\bf r};{\bf P})=G_z(\mathcal{R}^{T}(\gamma){\bf r};{\bf P}').\label{eq:G}
\end{equation}

Overall, with ${\bf P}=\{\gamma,{\bf P}'\}$, there are a total number of nine parameters ${\bf P}=\{\theta_1,\theta_2,\theta_3,x^0,y^0,z_x^0,z_y^0, w_{x}^0,w_{y}^0\}$ for a 2D Gaussian beam. To decompose a wavefront into $n_A$ beams as in Eq.~(\ref{eq:EG}) involves a total number of $11\times n_A$ real parameters $\{c_j,{\bf P}_j\}$. 


\subsection{Implementing the sample  constraints}\label{sec:SC}

For a single sample as in Fig.~\ref{Fig1} as well as multiple samples, our prior knowledge about the sample can often be  formulated into a list of cost-functions $C_{\mu}$ to constrain the expected 2D complex-valued phase shift distribution, $\varphi(z_A)=\varphi_A$. The constraints help to fix the phase-ambiguity associated with $E_{s,{\rm twin}}$ while locating the sample planes at $z=z_A$.  In the following we provide examples of such $C_{\mu}$-constraints, generalized for multiple samples.

\subsubsection{Finite spatial support} \label{sec:Ssupport}
The prior knowledge on the regime of interest where the atomic samples should be constrained within, as schematically illustrated in Fig.~\ref{Fig1}c with $\hat P_A$, is a condition commonly applied for coherent imaging~\cite{Koren1993,Miao1999,Sobol2014,Latychevskaia2019}. Here, for multiple atomic samples, the constraint can be formulated as a request to minimize the cost function,
\begin{equation}
  C_{\mu}=\sum_{A} \frac{1}{\mathcal{N}_A}\sum_{x,y} |\varphi(z_A)- \hat P_A \left(\varphi(z_A)\right)|^2.\label{eq:fsC}
\end{equation}
Here $\mathcal{N}_A= \sum_{x,y} |\hat P_A (\varphi(z_A))|^2$ normalizes the error signal from each sample at the respective $z=z_A$.

\subsubsection{Known phase angle}\label{sec:betaFix} 
As being outlined in Sec.~\ref{sec:intro} and discussed in Sec.~\ref{sec:cPhase2level}, for complex-valued atomic spectroscopy of a dilute gas, the phase angle $\beta={\rm arg}(\varphi_A)$ is often decided by the polarizability of single constituent atoms, $\beta_0={\rm arg}(\alpha)$, and is insensitive to transition saturation effects.  Therefore, the phase angle is often precisely known. The prior knowledge can be applied to constrain the $\langle E_s\rangle_G$-search, by minimizing 
\begin{equation}
  C_{\mu}=\sum_A \frac{1}{\mathcal{N}_A}\hat P_A \sum_{x,y}\left|{\rm Im}[\varphi(z_A)e^{-i\beta_0}]\right|^2.\label{eq:paC}
\end{equation}
That is, by unwrapping the known phase angle $\beta_0$ from the complex-valued phase shift, $\varphi_A e^{-i\beta_0}$ should be real, with vanishing imaginary component across each sample.  

The Eq.~(\ref{eq:paC})-minimization is a generalization to previous phase angle constraints in holographic microscopy, such as positive attenuation criterion~\cite{Latychevskaia2007} and phase object criterion~\cite{Smits2020}. Application of Eq.~(\ref{eq:paC}) requires the phase angle $\beta=\beta_0$ as precise prior knowledge. Practically, for free atoms with a line-center and linewidth not known precisely enough, then, $\beta_0(\Delta)$ can be fixed by exploiting the Eq.~\eqref{eq:varphi2level} relation across the atomic resonance under study first, during which the axial location $z_A$ for the standard samples are fixed too, as in Sec.~\ref{sec:zLoc}.


\subsubsection{Phase-angle uniformity}\label{sec:pUniform}
Without the precise $\beta_0$ knowledge, one still can exploit prior knowledge on $\beta$ in holography. A common situation is that the atomic sample is monomorphous~\cite{Turner2005}, with a uniform albeit not-precisely-known $\beta={\rm arg}(\varphi_A)$ across the sample. The condition can be formulated as to minimize
\begin{equation}
  C_{\mu}=\sum_A \frac{1}{\mathcal{N}_A}\hat P_A \left(|\sum_{x,y} \varphi(z_A)|-\sum_{x,y} |\varphi(z_A)|\right)^2.\label{eq:paU}
\end{equation}

\subsubsection{Other types of constraints}
Beyond the simple examples by Eqs.~(\ref{eq:fsC})(\ref{eq:paC})(\ref{eq:paU}), more sophisticated form of prior knowledge about the samples can be applied to tailor the cost functions $C_{\mu}(\varphi)$. For example, the finite support in Eq.~(\ref{eq:fsC}) can be replaced by a continuously varying $\hat P_A$ according to our expectation of the sample shape. Similar strategies can be applied to soften the $\beta_0$ bounds in Eq.~(\ref{eq:paC})~\cite{Latychevskaia2007}. Other cost functions may be tailored to encourage the $\varphi(z_A)$ distribution meeting our expectation of a mixed spatial-spectral characteristics, such as for atomic samples confined by a dipole trap~\cite{Babiker2013}.

\subsection{Direct constraints}\label{sec:direct}

\subsubsection{Direct constraints to Gaussian parameters}
Instead of formulating $C_{\mu}$ cost functions, simple knowledge about the sample can also be applied as direct constraints to the Gaussian parameters. For example, the finite spatial support in Eq.~(\ref{eq:fsC}) can be effectively enforced by limiting the range of ${\bf r}_A$ and $w_x(z_A),w_y(z_A)$ parameters (Eq.~(\ref{eq:G})). As another example, in the $|\varphi(z_A)|\ll 1$ weak scattering limit the phase-angle relation by Eq.~(\ref{eq:paU}) effectively enforce the wavefront radii of curvature $R_x(z_A), R_y(z_A)$ to follow the $\langle E_r\rangle_G$ values.

\subsubsection{Wavefront templates}

Direct constraints to groups of Gaussian parameters become particularly useful when the $E_s$ is expected to be characterized by certain wavefront template.  Examples include synthesizing the point-spread-functions of high-NA microscopes~\cite{LaRooij2023}, or for finely matching the expected shape of an interacting degenerate gas~\cite{Castin2001}. To form the template, the Gaussian-decomposition of the wavefront template is generally expressed as
\begin{equation}
  \tilde G_{\rm template}({\bf r})=\sum_g c_g G({\bf r}; {\bf P}_g).\label{eq:preG}
\end{equation}
The $\tilde G_{\rm template}$ with correlated Gaussian parameters $\{c_g,{\bf P}_g\}$ then enters the $\mathcal{L}_{\rm tot}$-minimization, together with other individual Gaussian beams if necessary. 

\subsubsection{Beyond paraxial}\label{sec:beyondP}
Finally, the Gaussian propagation by Eq.~(\ref{eq:Gw}) is based on paraxial approximation which limits the propagation distance $L$ to $L\ll (z^R_c)^2/\lambda$. To enhance the distance of faithful propagation,  any Gaussian beam that is tightly focused around $z=z_A$ can be expanded into a superposition of weakly focused Gaussians according to Eqs.~(\ref{eq:G})(\ref{eq:preG}).

\section{Out-of-focus effects}
\subsection{Defocused-twin regime}\label{sec:dftwin}

\begin{figure}[htbp]
\centering
\includegraphics[width=\linewidth]{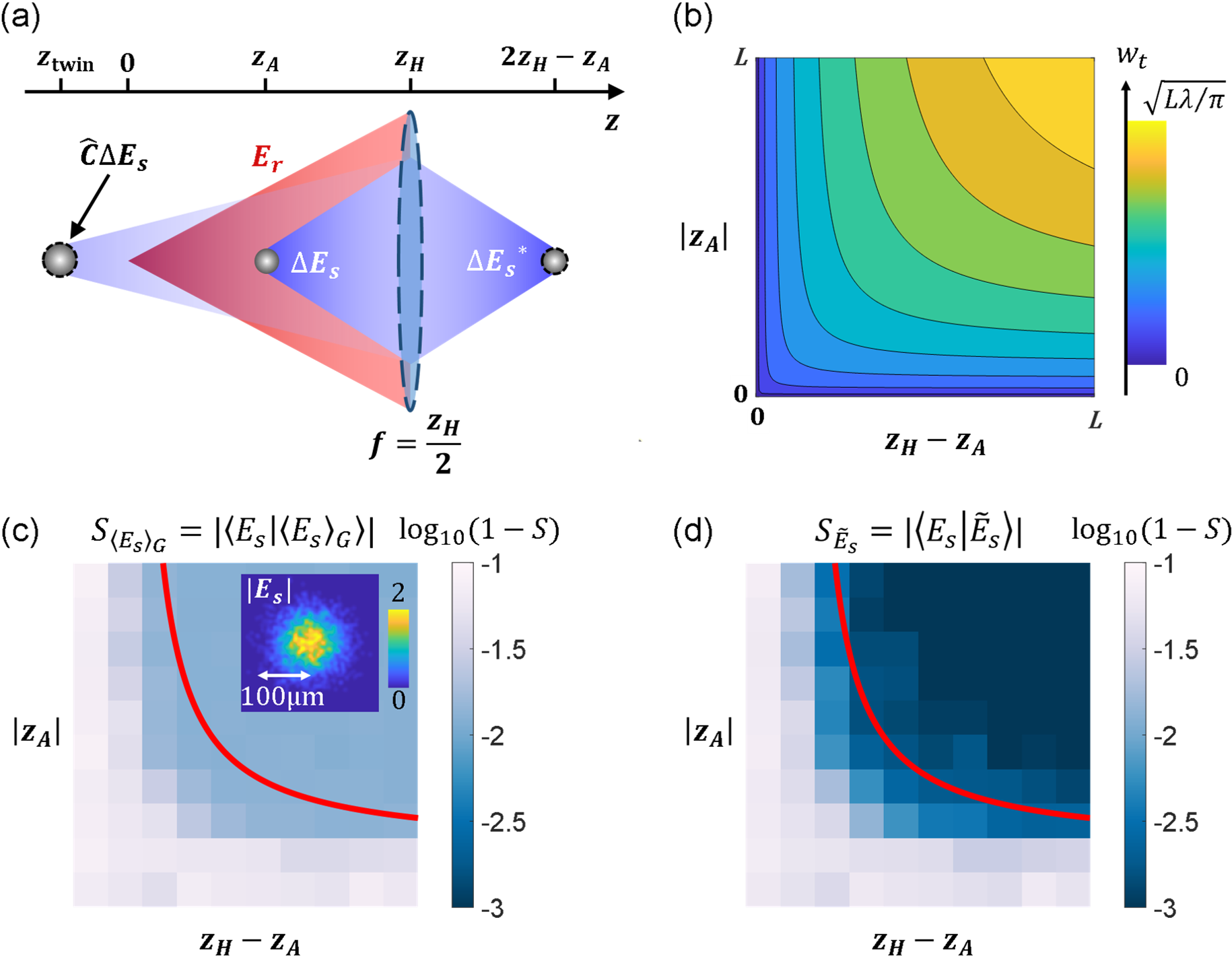}
 \caption{(a): Schematic diagram of the geometric relation between image and twin image when $z_A\textless z_H/2$. The ${\Delta E_s}^{*}$ is a mirror-image of $\Delta E_s$ at the other side of the holographic plane $z_H$. For spherical $E_r$, the $E_r/{E_r}^{*}$ term acts as a numerical lens with a focal length $f=z_H/2$. (b): The critical Gaussian beam size $w_t$ as a function of $z_A$ and $z_H-z_A$. (c): Fidelity of reconstructing $E_s$ using the Gaussian $E_G$ (Eq.~\eqref{eq:EG}) approximations (Eq.~\eqref{eq:inner}). Here the simulated $E_s$ follows a Gaussian profile with pixel-wise, Poissoinian atomic shot noise. The atom number is set to be 3500 in the simulation. 
The plot is averaged over 16 $E_s$ simulations. The red line sets the critical Gaussian $w_t$. A typical numerically generated $E_s$ from such a sample is given in the inset. (d): Fidelity plot similar to (c) but with  $\langle E_s\rangle_G$ replaced by $\tilde{E}_s$ (Eq.~\eqref{eq:EsApp}) to substantially improve the accuracy. 
}
\label{Fig6}
\end{figure}


In the Fig.~\ref{Fig1}a inline setup with the spherical $E_r$ illumination, the twin-image $E_{s,{\rm twin}}$ (Eq.~(\ref{eq:twin})) is paraxially focused to $z=z_{\rm twin}$. With $z_{\sigma}\ll z_A, z_H-z_A$, the $E_{s,{\rm twin}}$ is sufficiently defocused at $z=z_A$. In absence of imaging noise, isolation of $E_s$ from $E_{s,{\rm twin}}$ becomes possible by limiting the $E_s$ location $\hat P_A$ with tight enough size of order $\sigma$~\cite{Koren1993}. We refer this inline holography regime that supports simple twin-image removal as ``defocused-twin regime''. In the following we exploit Gaussian optics to derive a characteristic Gaussian width $w_t$, as a function of $z_A$ and $z_H-z_A$,
that sets the boundary of this regime for imaging small samples.

We consider that in the Fig.~\ref{Fig1}a setup the numerical reconstruction of certain $E_s$ is contaminated by an error term $\Delta E_s$, which is itself a Gaussian beam with complex radius of curvature  $q=\frac{1}{a+i b}$ at $z=z_A$. Within the paraxial approximation, the twin image, expressed as  $\Delta E_{s, \rm twin}=\Delta E_s^*E_r/E_r^*$ at the camera plane $z=z_H$, is also a Gaussian. We evaluate the twin profile $\Delta E_{s,\rm twin}$ at $z=z_A$ in four steps: propagation $\Delta E_s$ to $z_H$ plane; taking the complex conjugation; passing through an ideal thin lens with a focal length of $z_H/2$; and propagating to $z=z_A$ plane. The complex curvature of the $\Delta E_{s, \rm twin}$ at $z=z_A$ is given by:
\begin{equation}
    q_{\rm twin}=\frac{q(2z_A-z_H)+2z_A(z_H-z_A)}{2z_A-z_H-2q} \label{eq:qt}
\end{equation}
Now, if $|\Delta E_s|\ll |E_r|$, then it is easy to show that the artifact field $E_{\rm arti}=(\Delta E_s-\Delta E_s^*E_r/E_r^*)$ can be added freely to $E_s$ without affecting the expected hologram $I_e=|E_r+E_s|^2$. In other words, if the hologram data $(I,I')$ and $E_r$ are all the prior knowledge we have, then the $E_{\rm arti}$ term is free to contaminate $E_s$ during its reconstruction. However, if $\hat P_A$ has a width that is smaller than the smallest possible $E_{\rm arti}(z_A)$, then all the artifacts $E_{\rm arti}$ can be excluded. Here, the minimal width of $E_{\rm arti}(z_A)$
associated with the Gaussian $\Delta E_s$ and $\Delta E_{s,\rm twin}$ is set by equating their widths at $z=z_A$, that is, by equating the imaginary parts of $1/q$ and $1/q_{\rm twin}$. We obtain
\begin{equation}
    b^2=-a^2-\frac{2z_A-z_H}{z_A(z_H-z_A)}a+\frac{1}{z_A(z_H-z_A)}. \label{eq:b2}
\end{equation}
Next, by maximizing $b^2$ (see Eq.~\eqref{eq:Gw}), we obtain the minimal $E_{\rm arti}$ width
\begin{equation}
    w_t=\sqrt{\frac{2\lambda z_A(z_H-z_A)}{\pi z_H}}. \label{eq:wt}
\end{equation}

Typical $w_t$ according to Eq.~\eqref{eq:wt} is given in Fig.~\ref{Fig6}b.
With Eq.~\eqref{eq:wt}, the defocused-twin condition $z_{\sigma}\ll z_A, z_H-z_A$ is also expressed as $\sigma\ll w_t$. With $\sigma=\nu w_t$ and $\nu \ll 1$, the hologram $\delta I$ contains sufficient information to recover the Gaussian shaped $E_s$ without requiring additional prior knowledge. Furthermore, by converting $\langle E_s\rangle_G$ into its twin, $\langle E_{s,{\rm twin}} \rangle_G=\left(\frac{E_r}{E_r^*}\langle E_s\rangle_G^* \right)_{z=z_H}$, its contribution to $\delta I$ is subtracted away as in Eq.~(\ref{eq:halfstep}). The high-spatial-frequency twin-image residual spreads out at $z=z_A$ with a width that is at least $w_t/\nu$. Its impact to $E_s$ is therefore limited by $R \nu^2$ at least. Here $R\ll 1$ is the fractional residual of the Gaussian fit to $E_s$. As in the Fig.~\ref{Fig6}d example, for small enough samples ({\it e.g.}, $\nu \sim 0.3$) with good Gaussian-fits ({\it e.g.}, $R < 0.1$), the Eq.~(\ref{eq:EsApp}) approximation to $E_s$ can be highly accurate.

From Eq.~\eqref{eq:wt}, it is also clear that for a fixed $z_H$ distance between the point source and the camera plane,  $w_t$ reaches its maximum $\sqrt{\frac{\lambda z_H}{2\pi}}$ at $z_A=z_H/2$. In other words, for inline-holographic imaging of large samples under spherical wave illumination, one prefers to set $z_A\approx z_H/2$ to maximize the recordable $E_r$, $E_s$ wavefront mismatch for the efficient twin-removal.



\subsection{Sample sparsity}\label{sec:multipleA}

In Sec.~\ref{sec:principle}, our discussions of Gaussian-assisted holographic imaging exploit a single-sample example 
as illustrated in Fig.~\ref{Fig1}a. The procedure is then generalized in a straightforward manner to non-overlapping samples weakly displaced along $z$. Here, we more generally consider 3D distribution of samples. A necessary condition to separate $E_{s,A}$ from each sample is that the wavefronts from any two samples are orthogonal to each other, 
\begin{equation}
\langle E_{s,A}|E_{s,A'}\rangle=0.\label{eq:otho}
\end{equation}
Here the wavefront inner product is defined as
\begin{equation}
  \langle E_{a}|E_{b}\rangle \equiv \frac{\int {\rm d}x{\rm d}y E_{a}^*E_{b}}{\sqrt{(\int {\rm d}x{\rm d}y|E_{a}|^2)(\int {\rm d}x{\rm d}y|E_{b}|^2)}}.\label{eq:inner}
\end{equation}
It is easy to show that the inner product is invariant along $z$, and can therefore be evaluated at any $z-$plane. For the weakly displaced samples considered in Sec.~\ref{sec:principle} with $z_{A,A'}\equiv z_A-z_{A'}\ll z_{\sigma},z_{\sigma'}$, the Eq.~\eqref{eq:otho} orthogonality is guaranteed by non-overlapping samples within $z=z_A$, {\it i.e.}, $\sqrt{x_{A,A'}^2+y_{A,A'}^2}\gg \sigma+\sigma'$. Here $\sigma,\sigma'$ are the characteristic widths of the samples at $z=z_A,z_{A'}$ respectively.

On the other hand, when $E_{s,A}$, $E_{s,A'}$ have substantial overlap at either $z=z_A$ or $z=z_{A'}$, then the relative axial displacement must be large enough, $z_{A,A'}\gg z_{\sigma}+z_{\sigma'}$, to guarantee the wavefront orthogonality.

With Eq.~\eqref{eq:otho} and given sufficient additional prior knowledge about the samples, the Gaussian-decomposition of $E_s$ prescribed in Sec.~\ref{sec:principle} can be proceeded to obtain $\langle E_s\rangle_G=\sum_A\langle E_{s,A}\rangle_G$. Then, to approximately retrieve diffraction-limited $E_{s,A}$ using $\tilde E_s$ from Eq.~\eqref{eq:halfstep}, Eq.~\eqref{eq:EsApp} is modified as
\begin{equation}
  \begin{aligned}
    \tilde E_{s,A} &=\tilde E_s-\sum_{A'\neq A}\langle E_{s,A'}\rangle_G, \\
    E_{s,A}(z_A)&\approx\hat{P}_A( \tilde E_{s,A}(z_A)).\label{eq:EsAppb}
  \end{aligned}
  \end{equation}
Finally, Eq.~\eqref{eq:EsAppb} can be iteratively refined,  with the $\hat P_A$-support, by replacing the $\sum_{A'\neq A}\langle E_{s,A'}\rangle_G$ term in the first line with the resulting second line.

\section{Details of numerical implementation}
\subsection{Free propagation}
In the Fig.~\ref{Fig1}a setup, we assume the wavefronts propagate freely for $z>0$. The propagation operator $\hat U(\Delta z)$ is conveniently written in the angular spectrum basis as
\begin{equation}
        \hat U(\Delta z)=\mathcal{F}^{-1}e^{i\sqrt{k_0^2-k_x^2-k_y^2}\Delta z}\mathcal{F}.\label{eq:asm}
\end{equation}
Here $\mathcal{F}=\mathcal{F}_{x,y}^{k_x,k_y}$ represents 2D Fourier transformation such that $E(k_x,k_y,z)=\mathcal{F}_{x,y}^{k_x,k_y}(E(x,y,z))$. 

The free-space propagator is applied to retrieve $E_r^{(0)}$~\cite{Wang2022b} , as well as to obtain $E_s(z_A)$ with Eq.~(\ref{eq:halfstep}) during the final step of $\varphi(z_A)$ reconstruction.

\subsection{Nonlinear optimization}

We follow an adaptive moment estimation algorithm ~\cite{Kingma2015Adam} to minimize Eq.~(\ref{eq:loss}) in small steps. Taking advantage of the analytical Gaussian expression by Eq.~(\ref{eq:G}), we randomly sample a small fraction of $x,y$ pixels in $\delta I$ during the evaluation of $\mathcal{L}_H$. We find in most cases, the sparse sampling of about $1\%$ of full data is sufficient to ensure the convergence of the nonlinear optimization. Similarly, during the evaluation of $\mathcal{L}_A$ associated with {\it e.g.} Eqs.~(\ref{eq:fsC})(\ref{eq:paC})(\ref{eq:paU}), we find that by evaluating $5\%$ of pixels within $\hat{P}_A$ is sufficient to ensure the convergence.

To analyze a set of holograms from similar experimental measurements, the very initial guess of Gaussian parameters are obtained by fitting the Eq.~(\ref{eq:halfstep}) approximation to $E_s(z_A)$ but with $\langle E_s\rangle_G$ set to zero (the $E_s^{(0)}(\bar z_A)$ in Fig.~\ref{Fig2}b). Following the optimization for the first measurement, the optimal $z_A$, $\{c_j,{\bf P}_j\}$ parameters are transferred to initialize the optimization in similar measurements. Here, it is worth noting that while the parameter transfer greatly speed up the nonlinear optimization, the method is prone to cumulative numerical errors if the nonlinear optimization does not fully converge. To verify that the results are free from the accumulative errors, we often deliberately reset a set of parameters to far-from-optimal values during the parameter transfer.

\subsection{Aberration corrections}\label{sec:aberrApp}

Following Eq.~\eqref{eq:OTF1} and for the convenience of presentation, we define the ``ground-truth'' fields $E_{r,s}^{(0)}(x,y,z_A)=E_{r,s}(M x_V,M y_V,z_{V,A})$ between a specific pair of $z_A,z_{V,A}$ planes, with $M$ to be the magnification factor. With $U_{0,\rm img}$ to represent the perfect wavefront transformation, we have
\begin{equation}
    \begin{aligned} 
        E_r(z_A)&=U_{\rm img}  U_{0,\rm img}^{-1} E_{r}^{(0)}(z_A),\\
        E_s(z_A)&=U_{\rm img} U_{0,\rm img}^{-1} E_{s}^{(0)}(z_A).
    \end{aligned}\label{eq:OTF}
\end{equation}
The key insight about aplanatic aberration correction is that for $E^{(0)}(x_0,y_0,z_A)$ within a small enough imaging area at the $z_A$-plane, 
\begin{equation}
        E(x,y,z_A)= \int {\rm d}x_0 {\rm d}y_0 E^{(0)}(x_0,y_0,z_A) O(x-x_0,y-y_0).\label{eq:psf}
\end{equation}
Here $O(x,y)$ is the complex point-spread-function of our imaging system around ${\bf r}_A=(x_A,y_A,z_A)$. 

Experimentally, we obtain $O(x,y)$ by directly measuring $E_r'$, the probe beam $E_r$ (Fig.~\ref{Fig1}a) with its focus translated to the sample location ${\bf r}_A$. Similar to the $E_r$-phase recovery in Fig.~\ref{Fig1}b, we take intensity profiles of $E'_r$ at multiple planes to reconstruct the complex $E'_r$-profile, using the Gerchberg-Saxton iterations~\cite{Wang2022b}. By assuming $E_r'$ to be diffraction-limited itself, we obtain the complex point-spread function as $O(x,y)=E'_r/|E'_r|_{\rm max}$, up to the NA=0.3 numerical aperture~\cite{Alt2002}. Notice here $O(x,y)$ at $z=z_A$ can be propagated to other $z-$planes through Eq.~\eqref{eq:asm}, such as to obtain $O(x,y,z_H)$ at the $z=z_H$ camera plane to correct for the wavefronts there, as long as those wavefronts can be back-focused around ${\bf r}_A$. 

With $O(x,y)$ at hand, we exploit Eq.~\eqref{eq:psf} in two ways. First,  during Gaussian decomposition of $E_s$ for atomic sample centered at $x_A,y_A$, in particular for the evaluation of Eq.~\eqref{eq:lossH}, we simply multiply $O(x-x_A,y-y_A,z_H)$ to the associated $E_G$. In other words, at this stage we approximate the Gaussians profiles as 2-dimensional $\delta-$functions at $z=z_A$ and directly exploit Eq.\eqref{eq:psf} to estimate the aberrated Gaussians at the camera plane. Second, during  the final evaluation of $E_s$ (Eq.~\eqref{eq:EsApp}) and $E_r$ (Eq.~\eqref{eq:Ercorr}), we divide $E_s$ and a windowed $E_r$ by $O(k_x,k_y)$ in k-space. Here $E_r$ is selected to within an area-of-interest, with a size close to the Fig.~\ref{Fig2} display. To minimize edge effects, a Blackman-window is applied for the $E_r$ selection.

Finally, we note that in our imaging system with $M\neq 1$, the simple volumetric data process is subjected to the paraxial approximation, {\it i.e.}, by Taylor-expanding the $k_{x,y}$-dependent phase shift in Eq.\eqref{eq:asm} to the second order only. This paraxial condition is well-satisfied in the experimental demonstration where the axial displacements of samples are moderate (Fig.~\ref{Fig2}).

\section{On robust phase-angle spectroscopy}\label{sec:OBE}

\subsection{Justification of Eq.~\eqref{eq:varphi2level}}
The physics picture underlying the Eq.~\eqref{eq:varphi2level} approximation is detailed in Ref.~\cite{Zhao2022b}. In the following we illustrate the picture with a 2-level model. We then generalize the conclusion to an isolated hyperfine transition.

We consider the electric dipole operator $\hat {\bf d}={\bf d}e^{-i\omega t}+h.c.$ and a 2-level atom with ${\bf d}_{eg}$ as the dipole matrix element. The complex atomic dipole induced by the probe field ${\bf E}_r$ is expressed as $\langle {\bf d} \rangle = \rho_{ge}(t) {\bf d}_{eg}$~\cite{Steck2003}. The optical Bloch equation (OBE) for the $\rho_{eg}$ coherence reads
\begin{equation}
  i\dot \rho_{eg}= (-\Delta-i\Gamma/2)\rho_{eg}+\Omega_{eg}(\rho_{gg}-\rho_{ee}).\label{eq:reg}
\end{equation}
The Rabi frequency is defined as $\Omega_{eg}=-{\bf E}_r\cdot {\bf d}_{eg}/\hbar$. With $\rho_{eg}(0)=0$, Eq.~\eqref{eq:reg} leads to
\begin{equation}
  \rho_{eg}(t)= \frac{i}{2} \int_0^t \Omega_{eg}(\tau)(\rho_{gg}(\tau)-\rho_{ee}(\tau))e^{-i(\Delta+i\Gamma/2)(t-\tau)}{\rm d}\tau.\label{eq:reg2}
\end{equation}

The polarizability $\alpha$ sensed by holography, through $\varphi_A=\frac{1}{2}k_0 \rho_c \alpha$ according to Eq.~\eqref{eq:varphiR}, is expressed as~\cite{Zhao2022b}
\begin{equation}
\alpha = \frac{\int_0^{\tau_{\rm p}} {\bf E}^*_r \cdot \langle{\bf d}\rangle \, {\rm d}t}{\int_0^{\tau_{\rm p}} |{\bf E}_r|^2 \, {\rm d}t}.\label{eq:alpha}
\end{equation}
Following an integration by part for Eq.~\eqref{eq:reg2}, we have
\begin{equation}
  \int_0^{\tau_{\rm p}} {\bf E}^*_r \cdot \langle{\bf d}\rangle \, {\rm d}t \approx \int_0^{\tau_{\rm p}} \frac{\hbar|\Omega_{eg}(t)|^2}{\Delta+i\Gamma/2}(\rho_{ee}(t)-\rho_{gg}(t)){\rm d}t.\label{eq:alpha2}
  \end{equation}
We therefore find ${\rm arg}(\alpha)={\rm arg}(-\Delta+i\Gamma/2)$ to justify Eq.~\eqref{eq:varphi2level} for the 2-level atom. Importantly, for smooth pulse ${\bf E}_r(t)$ with $\tau_{\rm p}\gg 1/\Gamma$, the leading-order $1/(\Delta+i\Gamma/2)^2$ correction in the integrant of Eq.~\eqref{eq:alpha2} tends to average out itself ~\cite{Zhao2022b}. This makes Eq.~\eqref{eq:alpha2} highly accurate.

Extending the 2-level argument to an isolated hyperfine transition leverages the fact that the electric dipole coupling can be decomposed into a series of 2-level transitions, as discussed in Ref.~\cite{Shore2014}, for optical pulses with stationary polarization. Since the transitions between states $|g_m\rangle$ and $|e_m\rangle$, which are suitable superpositions of Zeeman sublevels, share the same detuning $\Delta$ and decay rate $\Gamma$, summing over these transitions on the right-hand side of Eq.~\eqref{eq:alpha2} results in a polarizability $\alpha$ that has the same phase angle as in the case of a 2-level atom.

To generalize Eq.~\eqref{eq:alpha2} to multi-level atoms as depicted in Fig.~\ref{FigSat}a requires all the $|g'\rangle\leftrightarrow|e\rangle$, $|g\rangle\leftrightarrow |e'\rangle$ and $|g'\rangle\leftrightarrow|e'\rangle$
transitions to be isolated from the target $|g\rangle\leftrightarrow|e\rangle$ transition in frequency~\cite{Zhao2022b}. In this case, when $|g\rangle\leftrightarrow|e\rangle$ is resonantly excited, we still have $\langle {\bf d}\rangle\approx \rho_{eg}{\bf d}_{ge}$ and furthermore Eq.~\eqref{eq:reg} remains approximately valid.

\subsection{Numerical model}
Our numerical model is based on vectorial light-atom interaction on a hyperfine manifold of alkaline~\cite{Sievers2015a,Qiu2022,Qiu2023}. The level structure is illustrated in Fig.~\ref{FigSat}a. For a specific probe ${\bf E}_r(t)$ form, with the atomic state initiated as $\rho(0)$, we integrate the optical Bloch equation
\begin{equation}
    i\hbar\dot\rho=H_{\rm eff}\rho-\rho H_{\rm eff}^{\dagger}+\hbar\sum_{l}C^l\rho C^{l \dagger}\label{eq:OBE}
\end{equation}
to obtain the time-dependent $\rho(t)$ between $0<t<\tau_{\rm p}$. The atomic polarizability $\alpha$ is then evaluated according to Eq.~\eqref{eq:alpha}, with which $\varphi_A$ is modeled in the Beer-Lambert limit according to Eq.~\eqref{eq:varphiR}.


The effective, non-Hermitian Hamiltonian is
    \begin{equation}
        \begin{split}
            H_{\rm eff} = &\hbar\sum_{a=e,e'}\qty(\omega_{e}-\omega_{e0}-i\Gamma/2)\sigma^{a_ma_m} + \\
            &\hbar\sum_{b=g,g'}\qty(\omega_{c}-\omega_{g0})\sigma^{b_nb_n}+\\
            & \frac{\hbar}{2}\sum_l \sum_{a={e,e'}}\sum_{b=g,g'}\Omega_{a_m b_n}^l(t)\sigma^{a_m b_n} + {\rm h.c.}
        \end{split}\label{eq:HD1}
    \end{equation}
The $m,n$ sub-indices in $a_m,b_n$ label the magnetic quantum number of the Zeeman sublevels, for which we follow the Einstein convention to implicitly sum over if repeating.  The  $\sigma^{a_m b_n} = \op{a_m}{b_n}$, $\sigma^{b_n a_m } = \op{b_n}{a_m}$ are the raising and lowering operators between states  $|a_m\rangle$ and $|b_n\rangle$. Similar $\sigma$ operators are defined for all the other $|a_m\rangle$, $|b_n\rangle$ state combinations. The $\omega_{e0}, \omega_{g0}$ are decided by the energy of reference level in the excited and ground state manifolds respectively, chosen as the top hyperfine levels in our case. The probe pulse Rabi frequency,
    \begin{equation}
        \begin{split}
            \Omega_{a_mb_n}^l (t)&\equiv -\frac{\mel{a_m}{\vb{d}_l \cdot {\bf E}_r(t)}{b_n}}{\hbar},
        \end{split}\label{eq:Rabi}
    \end{equation}
is accordingly written in the $\omega_{e0,g0}$ frame under the rotating wave approximation. The ${\bf d}_l$ with $l=-1,0,1$ are projections of the electric dipole operator to $\{{\bf e}_-,{\bf e}_z, {\bf e}_+\}$ directions, respectively. 

  
Finally, the quantum jump operators associated with spontaneous emissions are given by 
\begin{equation}
    C^l=\sqrt{\Gamma}\sum_{a=e,e'}\sum_{b=g,g'} \mathcal{C}_{ b_n a_m}^l \sigma^{b_n a_m},
\end{equation}
with $l=-1,0,1$, for the $\sigma^+,\pi,\sigma^-$ transitions. We accordingly have $\sum_l\mathcal{C}^{l\dagger}\mathcal{C}^l=\Gamma \sum_{a=e,e'} \sigma^{a_m a_m}$.

\subsection{Details on Fig.~\ref{FigSat} simulation}\label{sec:Fig2detail}
The $|g\rangle-|e\rangle$ transition in Fig.~\ref{FigSat}a corresponds to the $F=2-F'=3$ hyperfine D2 transition of $^{87}$Rb. The transition properties in the simulation follow Ref.~\cite{Steck2003}. To produce the Fig.~\ref{FigSat}(b-d) data we evaluate $\alpha$ according to Eq.~\eqref{eq:alpha} for $\tau_{\rm p}=10~\mu$s. The saturation parameter $s_0$ is chosen according to the desired $\Omega_{\rm p}$. Specifically, with the Clebsch-Gordan coefficients $\mathcal{C}_{a_n b_m}^{l}$, the Rabi frequencies in Eq.~(\ref{eq:Rabi}) are rewritten as $\Omega_{a_m b_n}^l(t)= \Omega_{\rm p}(t)\mathcal{C}_{a_m b_n}^l/\eta$, with $\Omega_{\rm p}=\sqrt{s_0/2}\Gamma$ as in Fig.~\ref{FigSat}a. The $\eta=\sqrt{I_s^{(0)}/I_s}$ factor accounts for polarization-dependent interaction strength variation (Eq.~\eqref{eq:varphi2level}) that is included into the $s_0\equiv I/I_s$ evaluation. 
The $I_s^{(0)}$ is the saturation intensity for the hyperfine cycling transition driven by circularly polarized light, $I_s^{(0)}=1.67~{\rm mW/cm}^2$ here. The  $I_s$ is instead evaluated for the specific light polarization and atomic state under consideration, $I_s=3.05~$mW/cm$^2$ for the $\pi$ polarized light  according to the steady-state atomic response~\cite{Steck2003}.

We consider square-shaped probe pulse $E_r$ as depicted in Fig.~\ref{FigSat}a, with rising and falling edges sine-smoothed within 200~ns to mimic the actual probe pulse controlled by acousto-optical modulation. With $\tau_{\rm p}=10~\mu$s$\gg 1/\Gamma$, we find that the actual edge-shapes  hardly affect our numerical observations. To integrate Eq.~\eqref{eq:OBE}, we set the atomic initial state to equally populate the Zeeman sublevels of 5$S_{1/2}$, $F=2$, denoted as $|g\rangle$ in Fig.~\ref{FigSat}a. The probe polarization is chosen along ${\bf e}_y$ which is also set as the quantization axis.  However, we numerically verified that details on the Zeeman population  and the (constant) probe polarization only affect $|\alpha|$ and therefore $|\bar{\varphi}|$, instead of the phase angle $\beta$. This insensitivity to the Zeeman population nor coherence is consistent with the observation that the phase-angle spectroscopy is robust against optical pumping and alignments~\cite{Happer1972} during the prolonged strong probe, as in the experiments (Sec.~\ref{sec:exp}).


The $F = 2 - F' = 3$ transition, denoted as $|g\rangle - |e\rangle$ in Fig.~\ref{FigSat}a, is a closed transition. This transition allows the atom to reach a steady state when excited by a prolonged probe with $\tau_{\rm p} \gg 1/\Gamma$, leading to $N_s \propto s / (1 + s)^2 \Gamma \tau_{\rm p}$~\cite{Sobol2014a}, which improves with increasing probe duration $\tau_{\rm p}$. In contrast, for open transitions, the atomic population decreases under a strong and extended probe. 
A systematic study on the excitation bandwidth dependence of the Eq.~\eqref{eq:alpha} approximation, particularly for open and closed transitions and on the influence of nearby levels, will be presented in a future publication~\cite{foot:future}.

The power-broadening effects in ${\rm OD}$ and $\phi$, as well as their suppression in $\beta$, are clearly illustrated in Fig.~\ref{FigSat}(b-d). By fitting the ${\rm OD}$ data with Lorentzian, we obtain linewidths of $\tilde{\Gamma} = 2\pi \times (6.2, 8.4, 14.5)~\text{MHz}$ for $s_0 = 0.1, 1, 5$, respectively. In contrast, fitting ${\rm tan}\beta = (\delta-\Delta) / (2\tilde{\Gamma})$, weighted by $|\varphi_A|$, consistently yields $\tilde{\Gamma} = 2\pi \times 6.1~\text{MHz}$, which is in agreement with the input value of $\Gamma = 2\pi \times 6.1~\text{MHz}$. Since the $F' = 2, 3$ hyperfine levels are well separated, the line-center shift $\delta$ is also small, about 7~\text{kHz} and 21~\text{kHz} at $s_0 =1, 5$ respectively. These findings are consistent with our experimental observations (Fig.~\ref{Fig4}, Fig.~\ref{Fig8}).


\subsection{Radiation pressure and Doppler effect}\label{sec:Doppler}
So far the analysis in this section ignored the center-of-mass  motion. The atomic motion with velocity ${\bf v}$ leads to Doppler shift of the transition frequency, $\omega_{eg}\rightarrow\omega_{eg}+{\bf k}_0\cdot {\bf v}$. Here ${\bf k}_0$ is the phase gradient of the probe $E_r({\bf r})$ locally seen by the atom. Laser cooling~\cite{MetcalfBook} substantially reduces ${\bf v}$ to improve the atomic spectroscopic precision~\cite{Li2020}. For example, with the $20~\mu$K atomic sample in this work, the atomic motion merely leads to $100~$kHz inhomogeneous broadening, undetectable for us in presence of the $\Gamma=2\pi\times 6$~MHz natural linewidth background.  On the other hand, for a strong and elongated probe, the atomic motion is accelerated by radiation pressure during the measurement, which may lead to line distortion as we observed experimentally in Fig.~\ref{Fig4}. 

We estimate the Doppler bias associated with the radiation pressure~\cite{MetcalfBook} for a moderate period of probe time $\tau_{\rm p}$, during which $\delta \omega_{eg}\ll \Gamma$ is not substantial to affect the optical excitation efficiency. The estimation is for a closed transition as in Fig.~\ref{FigSat}a so that $\rho_{ee}\approx \frac{s}{2 (1+s)}$ is reached in $\sim 1/\Gamma$ time scale. The average Doppler shift is given by $\delta \omega_{eg}=k_0\delta v/2$, with $\delta v=\frac{s}{2(1+s)}\Gamma \tau_{\rm p} v_r$ to be the final velocity boost, and $v_r=\hbar k_0/m$ is the atomic recoil velocity. We therefore have
\begin{equation}
  \delta\omega_{eg}\approx \frac{s_0 \Gamma^2}{4\Delta^2+(1+s_0)\Gamma^2}\frac{\Gamma}{2}\omega_r \tau_{\rm p}.\label{eq:Doppler}
\end{equation}
Here $\omega_r=\hbar k_0^2/2m$ is the recoil frequency, for example, $\omega_r=2\pi\times 3.771~$kHz for $^{87}$Rb D2 line in this work~\cite{Steck2003}.  

We note the Doppler shift generally exists in atomic spectroscopy with free-flying atoms~\cite{Brown2013a}. For heavy alkaline with $\omega_r\ll \Gamma$, such as rubidium in this work, the effect can be mitigated by choosing $\omega_r \tau_{\rm p}\ll 1$, even at large $s_0$. Then, multiple probes can be interleaved with laser cooling pulses to effectively increase $\tau_{\rm p}$ without increasing the $\delta \omega_{eg}$ bias. This is demonstrated in ref.~\cite{Zhao2022b} where we also showed that the interleaved cooling and trapping help to suppress atomic diffusion to keep the spatial resolution of tightly confined samples. An ultimate way to eliminate the radiation-pressure systematic associated with Eq.~\eqref{eq:Doppler} is to tightly confine the laser-cooled atoms to a lattice~\cite{Marti2018}. 

The Doppler shift can also be corrected using the model in Eq.~\eqref{eq:Doppler}, as being effectively demonstrated with Fig.~\ref{Fig4}e. However, it is important to note that Eq.~\eqref{eq:Doppler} does not account for collective dipolar interactions, which can substantially modify the radiation pressure received by samples at high optical depth (OD)~\cite{Bachelard2016}, as shown in Fig.~\ref{Fig4}i. In this work, with both ${\rm OD}$ and $\phi$ resolved through complex spectroscopic imaging, we anticipate that the holographic technique demonstrated here will aid in refining the Eq.~\eqref{eq:Doppler} correction with the OD-knowledge to better suppress the radiation-pressure-induced systematic shift $\delta \omega_{eg}$ in atomic spectroscopy.


\section{Imaging larger samples}\label{sec:large}

\begin{figure}[htbp]
  \centering
  \includegraphics[width=\linewidth]{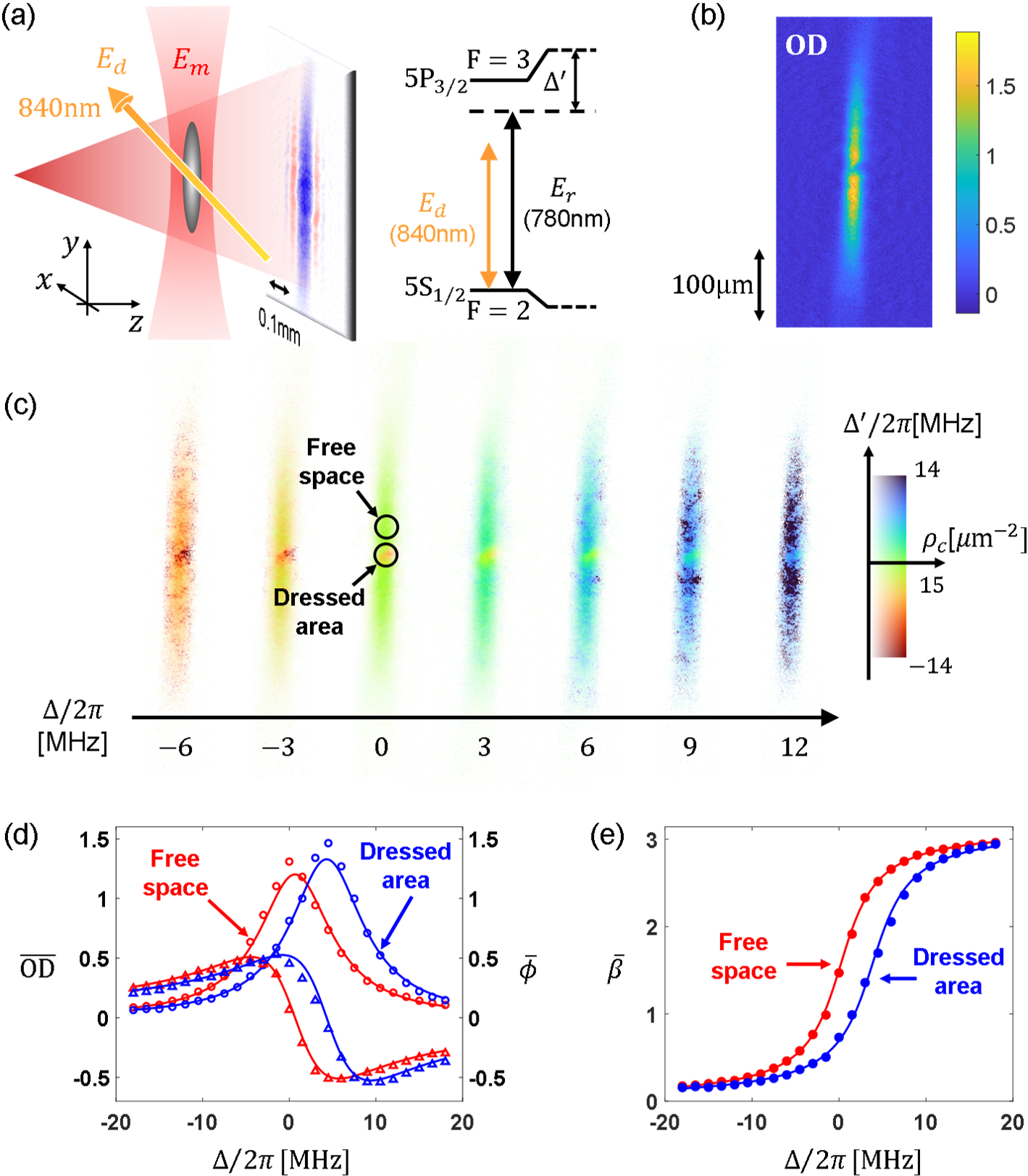}
   \caption{Spectroscopic imaging of a cigar-shaped atomic sample, released from a dipole trap ($E_m$), and subjected to a ``dimple'' $E_d$ dressing.  (a): The Experimental setup and the level diagram. (b): Typical optical depth image, processed from $\delta I$ averaged over 9 exposures to enhance the display. Here the probe detuning is $\Delta/2\pi=0~\rm{MHz}$. (c): $(\Delta',\rho_c)$ in color domain plots. Here $\rho_c$ is column density, and  $\Delta'=-\frac{\Gamma}{2} {\rm tan}\beta $ is the local detuning, both derived from the complex $\varphi_A$-data similar to Fig.~\ref{Fig4}a. (d): ${\rm ROI}-$ averaged $\overline{\rm{OD}}$ and $\overline{\phi}$ spectroscopy. The two Region-Of-Interest (ROI) are marked in (c). The red (blue) curves correspond to region without (with) the $E_d$-dressing. The solid lines are according to Eq.~\eqref{eq:varphi2level} with $\xi(s)\propto 1/(1+s)$. (e): The  phase-angle spectroscopy of the same data. Similar to Fig.~\ref{Fig4}, the fit for the free-space data with ${\rm tan}\beta=-2\Delta/\tilde \Gamma$ here gives $\tilde \Gamma=2\pi\times 6.1~$MHz to agree with $\Gamma$, in contrast to the substantial power-broadening in (d).}
\label{Fig8}
 \end{figure}

The experimental work presented in  Sec.~\ref{sec:exp} exploits small samples. Due to their regular shapes, a total number of $n_A=3$ Gaussians suffices the decomposition of $E_{s,A}$ for each sample. In this section, we provide an example of imaging larger, more complex samples. 

As in Fig.~\ref{Fig8}a, the cigar-shaped sample is released from the 1064~nm dipole trap (which acts as the $E_m$-field in the Fig.~\ref{Fig5} measurement). During the $\tau_{\rm p}=20~\mu$s exposure, a 30~mW ``dimple trap'' $E_d$ at $840~$nm is switched on to dress the atoms
with its $w\approx 10~\mu$m focal spot. Similar to the Sec.~\ref{sec:exp} experiment, we repeat the atomic sample preparation to take the $\{I,I'\}$ data-set with $\Delta/2\pi$ scanned between $-$18~MHz and $+$18~MHz. Part of the data are presented in Fig.~\ref{Fig8}. The {\rm OD}-image in Fig.~\ref{Fig8}b and the $(\Delta',\rho_c)$- images in Fig.~\ref{Fig8}c are all derived from the complex $\varphi_A$-data reconstructed from ``single-shot'' $\delta I$, as following. Similar to Fig.~\ref{Fig2}, here each $\delta I$ is also averaged over nine repetitions to enhance the display. We follow the Sec.~\ref{sec:principle} prescription with $n_A=16$ for the Gaussian decomposition of $E_s$. There are eight Gaussians to capture the global cigar shape, and another eight Gaussians to capture the local structure around the dimple-dressing beam. We then obtain $\tilde E_s$ according to Eq.~\eqref{eq:halfstep} and approximated $\varphi_A$ according to Eq.~\eqref{eq:EsApp}. In Fig.~\ref{Fig8}c at different detuning $\Delta$, we consistently see the local Stark-shift spot with $\delta \omega_{eg}\approx 2\pi\times 4~$MHz. The value is again consistent with expectation at the dressing laser power. 

Similar to Fig.~\ref{Fig4}, we average the $\varphi_A$-data (not shown) over two circular Region-Of-Interest(ROI), suggested in the Fig.~\ref{Fig8}c $(\Delta',\rho_c)$ plot, to compile $\overline{\rm OD}$, $\bar{\phi}$ in Fig.~\ref{Fig8}(d) and $\bar\beta$ data in Fig.~\ref{Fig8}(e). By fitting Fig.~\ref{Fig8}(d) data for free-space atoms with Eq.~\eqref{eq:varphi2level}, the probe intensity parameter $s_0\approx 1.2$ and the transition linewidth $\tilde \Gamma=2\pi\times 6.0~$MHz$~\approx \Gamma$ are extracted, within expectation  (For comparison, simple Lorentzian fits to $\overline{\rm OD}$ lead to power-broadened linewidth of about 8.5~MHz.). The fits for the phase-angle in Fig.~\ref{Fig8}e instead directly apply the ${\rm tan}\beta=-2\Delta/\tilde \Gamma$ relation, same as for Fig.~\ref{Fig4}(d)(h), to arrive at $\tilde \Gamma=2\pi\times 6.1~$MHz for the free-space atoms. For the Fig.~\ref{Fig8}c $(\Delta',\rho_c)$-images, the local detuning $\Delta'=-\frac{\Gamma}{2}{\rm tan}\beta$ is derived from $\beta={\rm arg}[\varphi_A]$. The atomic column density $\rho_c$ is estimated with Eq.~\eqref{eq:varphi2level} with $\xi(s)=\eta^2/(1+s)$ (Appendix~\ref{sec:Fig2detail})~\cite{Steck2003}. 


\pagebreak

\bibliography{SG}

\begin{thebibliography}{114}
\bibitem [{foo({\natexlab{a}})}]{foot:abre}%
  \BibitemOpen
  \bibinfo {note} {For the notation conciseness, we may omit the $x,y$ variables, or $x,y,z$ variables, when expressing the complex and real fields.}\BibitemShut {Stop}%
\bibitem [{foo({\natexlab{c}})}]{foot:future}%
  \BibitemOpen
  \bibinfo {note} {A detailed numerical study on the phase angle robustness will be given in a future publication.}\BibitemShut {Stop}%
\bibitem [{foo({\natexlab{d}})}]{foot:LFNB}%
  \BibitemOpen
  \bibinfo {note} {The accuracy of the saturation spectroscopy lock is limited by systematic offset associated with ambient fields and beam alignments. We therefore calibrate the offset with standard absorption imaging with cold atoms. The $\Delta$ values presented in Figs.~\ref{Fig4}~\ref{Fig5}~\ref{Fig8} are according to the fixed offset for all the measurements.}\BibitemShut {Stop}%
\bibitem [{foo({\natexlab{e}})}]{foot:LFN}%
  \BibitemOpen
  \bibinfo {note} {To generate error signal the laser diode current is weakly dithered at 250~kHz. The periodic frequency modulation, with an amplitude below 1~MHz, is efficiently averaged out during the $\tau_{\rm p}=20~\mu$s probe time. By observing the error signal, we estimate that the deviation of center frequency of our probe laser is below 100~kHz when the laser is frequency-locked.}\BibitemShut {Stop}%
\bibitem [{foo({\natexlab{f}})}]{foot:exc}%
  \BibitemOpen
  \bibinfo {note} {By excluding the $|\Delta|<4~$MHz data, which is more prone to laser intensity fluctuations due to the Doppler bias, $\Delta\bar\beta$ improves to $0.11$ and $0.031$ for the $\Circled{1}$ and $\Circled{4}$ samples respectively.}\BibitemShut {Stop}%
\bibitem [{foo({\natexlab{g}})}]{foot:LFN2}%
  \BibitemOpen
  \bibinfo {note} {By comparing the shot-to-shot data between different samples in the Fig.~\ref{Fig4} data, noticeable $\bar\beta-$correlations can be found. The relative phase-angle variations are closer to the photon-shot-noise limit set by Eq.~\eqref{eq:db}.}\BibitemShut {Stop}%
\end{thebibliography}
\bibliographystyle{revTex2}
\end{document}